\documentclass[useAMS,usedcolumn,usenatbib,usegraphicx]{mn2e}

\usepackage{lineno}

\pdfoutput=1
\usepackage{times}  

\usepackage{listings}
\usepackage{graphics}
\usepackage{subfigure}
\usepackage{graphicx}
\usepackage{amssymb}
\usepackage{aas_macros}
\usepackage{verbatim}
\usepackage{color}
\usepackage{url}
\usepackage[dvipsnames]{xcolor}
\usepackage{booktabs}

\usepackage{eso-pic}

\AddToShipoutPictureBG*{%
  \AtPageUpperLeft{%
    \hspace{17.3cm}%
    \raisebox{-3.5\baselineskip}{%
      \makebox[0pt][l]{\textnormal{DES 2015-0085}}
}}}%

\AddToShipoutPictureBG*{%
  \AtPageUpperLeft{%
    \hspace{15.5cm}%
    \raisebox{-4.5\baselineskip}{%
      \makebox[0pt][l]{\textnormal{FERMILAB-PUB-16-003-AE}}
}}}%

\long\def\symbolfootnote[#1]#2{\begingroup%
\def\thefootnote{\fnsymbol{footnote}}\footnote[#1]{#2}\endgroup}

\newcommand{\band}[1]{{\rm #1}}

\usepackage{xspace}

\newcommand{\eg}{e.g.\xspace}

\newcommand{\vs}{vs.\xspace}

\mathchardef\mhyphen="2D

\newcommand{\roughly}{\ensuremath{ {\sim}\,} }

\newcommand{\unit}[1]{\ensuremath{\mathrm{\,#1}}\xspace}

\newcommand{\km}{\unit{km}}
\newcommand{\pc}{\unit{pc}}
\newcommand{\kpc}{\unit{kpc}}
\newcommand{\second}{\unit{s}}

\newcommand{\Msolar}{\ensuremath{M_\odot}}
\newcommand{\Msun}{\ensuremath{M_\odot}}

\newcommand{\yr}{\unit{yr}}
\newcommand{\Myr}{\unit{Myr}}
\newcommand{\Gyr}{\unit{Gyr}}
\newcommand{\mas}{\unit{mas}}

\newcommand{\figref}[1]{Figure~\ref{fig:#1}}

\newcommand{\code}[1]{\texttt{#1}\xspace}
\newcommand{\Trilegal}{\code{Trilegal}}

\providecommand\physrep{\ref@jnl{Phys.~Rep.}}%
\providecommand\apjs{\ref@jnl{ApJS}}%
\providecommand{\jcap}{\ref@jnl{JCAP}}%

\newcommand{\colourp}[2]{\ensuremath{(\band{#1} - \band{#2})}\xspace}

\newcommand{\omegacen}{Omega Centauri\xspace}
\newcommand{\ra}{\ensuremath{\mathrm{RA}}\xspace}
\newcommand{\dec}{\ensuremath{\mathrm{DEC}}\xspace}

\newcommand{\beq}{\begin{equation}}
\newcommand{\eeq}{\end{equation}}
\newcommand{\beqa}{\begin{eqnarray}}
\newcommand{\eeqa}{\end{eqnarray}}

\title[DES: more than Dark Energy]{The Dark Energy Survey: more than dark energy  - an overview}

\author[The Dark Energy Survey Collaboration]{
\Large
\parbox{\textwidth}{
Dark Energy Survey Collaboration:
T.~Abbott$^{1}$,
F.~B.~Abdalla$^{2}$,
J.~Aleksi\'c$^{47}$,
S.~Allam$^{3}$,
A.~Amara$^{4}$,
D.~Bacon$^{6}$,
E.~Balbinot$^{46}$,
M.~Banerji$^{7,8}$,
K.~Bechtol$^{56,57}$,
A.~Benoit-L{\'e}vy$^{13,2,12}$
% ...
\\
}
\vspace{0.2cm}\\ 
\Large $\star$ Corresponding author: \texttt{\rm \texttt{o.lahav@ucl.ac.uk}}}

\begin{document}


\date{Accepted 2015 ???. Received 2015 ???; in original form 2015 ???}

\pagerange{\pageref{firstpage}--\pageref{lastpage}} \pubyear{2002}

\maketitle

\label{firstpage}

\begin{abstract}
This overview article describes the legacy prospect and discovery potential of the Dark Energy Survey (DES) beyond 
cosmological studies, illustrating it with examples from the DES early data. 
DES  is using a wide-field camera (DECam) on the 4m Blanco Telescope in Chile to image 5000 sq deg of the sky in five filters 
({\it grizY}). By its completion the survey is expected to have 
generated a catalogue of 300 million galaxies with photometric redshifts and 100 million stars.
In addition, a time-domain survey search over 27 sq deg is expected to yield a sample of 
thousands of Type Ia supernovae and other transients. The main goals of DES 
 are to characterise dark energy and dark matter, and to
  test alternative models of gravity;  these goals will be pursued by studying large scale structure,
  cluster counts, weak gravitational lensing  and Type Ia supernovae.
 However, DES also provides a rich data set which allows us
  to study many other aspects of astrophysics. In this paper we focus
  on additional  science with DES, emphasizing areas where the
  survey makes a difference with respect to other current surveys. The
  paper illustrates, using early data (from `Science
  Verification', and from the first, second and third seasons of observations), what DES can tell us about the
  solar system, the Milky Way, galaxy evolution, quasars, and other topics. In
  addition, we show that if the cosmological model is assumed to be
  $\Lambda$ + Cold Dark Matter (LCDM) then important astrophysics can be
  deduced from the primary DES probes. Highlights from DES early data
  include the discovery of 34 Trans Neptunian Objects, 17 dwarf satellites of
  the Milky Way, one published  $ z >  6$  quasar (and more confirmed) 
  and two published superluminous
  supernovae (and more confirmed).

\end{abstract}



\section{Introduction}

Many ongoing and planned imaging and spectroscopic surveys  aim at measuring dark energy (DE)  and other cosmological parameters. However, these surveys map huge volumes of the cosmos and they can be used to study many other astrophysical  objects and phenomena. Here we focus on the Dark Energy Survey (DES)\footnote{\url{http://www.darkenergysurvey.org/}}, illustrating its power with the early `Science Verification' (SV)\footnote{For public data release see \url{http://des.ncsa.illinois.edu/releases/sva1}}, first year (Y1),  second year (Y2) and third year (Y3)  data. We consider non-cosmology science derived in two ways: (i) science outside the original scope of DES: the solar system, the Milky Way, galaxy evolution, quasars (QSOs) , and transients; and (ii) astrophysics resulting from cosmology probes if the cosmology is already assumed, say to be $\Lambda$+Cold Dark Matter (LCDM). The motivation here is to emphasize where DES data make a difference with respect to current data, rather than to cover everything one can do with DES. This paper illustrates science with DES alone, as well as in combination with other data sets, methods used, and DES applications for the future.
This paper aims to present DES to a wide readership.
First, for those who will be using data from the DES public archive, this paper demonstrates what can be done with the public data.\footnote{
In addition to raw survey images released through the NOAO Science Archive (NSA) twelve months after they are taken, reduced, calibrated images processed through the DES Data Management (DESDM) System are being made publicly available through the NSA. In addition, DES will make two public releases of co-added images and catalogs produced and served by DESDM, one based on the first two seasons of data, the second based on the full survey data set.}
Second, this paper targets those who are applying for community observing time on the Dark Energy Camera (DECam), in providing ideas about the system's capabilities. Third, this paper could be of interest to those conducting current and future imaging and spectroscopic surveys,
e.g., the  Hyper Suprime Cam (HSC)\footnote{\url {http://www.naoj.org/Projects/HSC/}}, the Kilo-Degree Survey (KiDS)\footnote{\url{http://kids.strw.leidenuniv.nl/}},  the Large Synoptic Survey Telescope (LSST)\footnote{\url{ http://www.lsst.org/}}, \emph{Euclid}\footnote{\url{http://www.euclid-ec.org/}}, the Wide-Field Infrared Survey Telescope (WFIRST), the Subaru Prime Focus Spectrogrph (PFS)\footnote{\url{http://pfs.ipmu.jp/factsheet/}}, the Dark Energy Spectroscopic Instrument (DESI)\footnote{\url{http://desi.lbl.gov/}} and 4MOST\footnote{\url{http://www.4most.eu/}}.

 \cite{White2007} made the point that there is a  risk that DE-motivated projects may focus too strongly 
on a single science question. 
This paper is partly intended to demonstrate that 
even if a project is optimised for DE searches, it can and does have numerous spin-offs, including some well beyond cosmology.
It is anticipated that the next generation of DE-inspired projects (eg. LSST, DESI, Euclid, WFIRST) will also have such legacy value 
\footnote{A companion DES overview paper (in preparation) will describe the survey and the DES project in more detail, as well as cosmological studies 
such as Dark Energy, Dark Matter and other cosmological properties,
 e.g. cosmological parameters from early DES weak lensing data 
\citep{des_ta_2015} 
 and forecast for neutrino mass from DES large scale structure \citep{lahav2010}.}.

The outline of this paper is as follows. Section 2 summarizes the DES survey, its properties, and an inventory of objects recorded so far and expected from the full survey. 
Sections 3 to 11, are ordered by the proximity of objects to the Earth,  and cover the solar system, the Milky Way, galaxy evolution (including galaxy biasing and  intrinsic alignments), clusters, strong lensing, quasars, supernovae, and the follow-up of gravitational-wave alerts.
The last section summarises the results, and forecasts  what legacy science DES is likely to yield after completing its five seasons. 



\section[]{The Dark Energy Survey}




\begin{figure*}
  \centering
 \includegraphics[height=0.85\textheight]{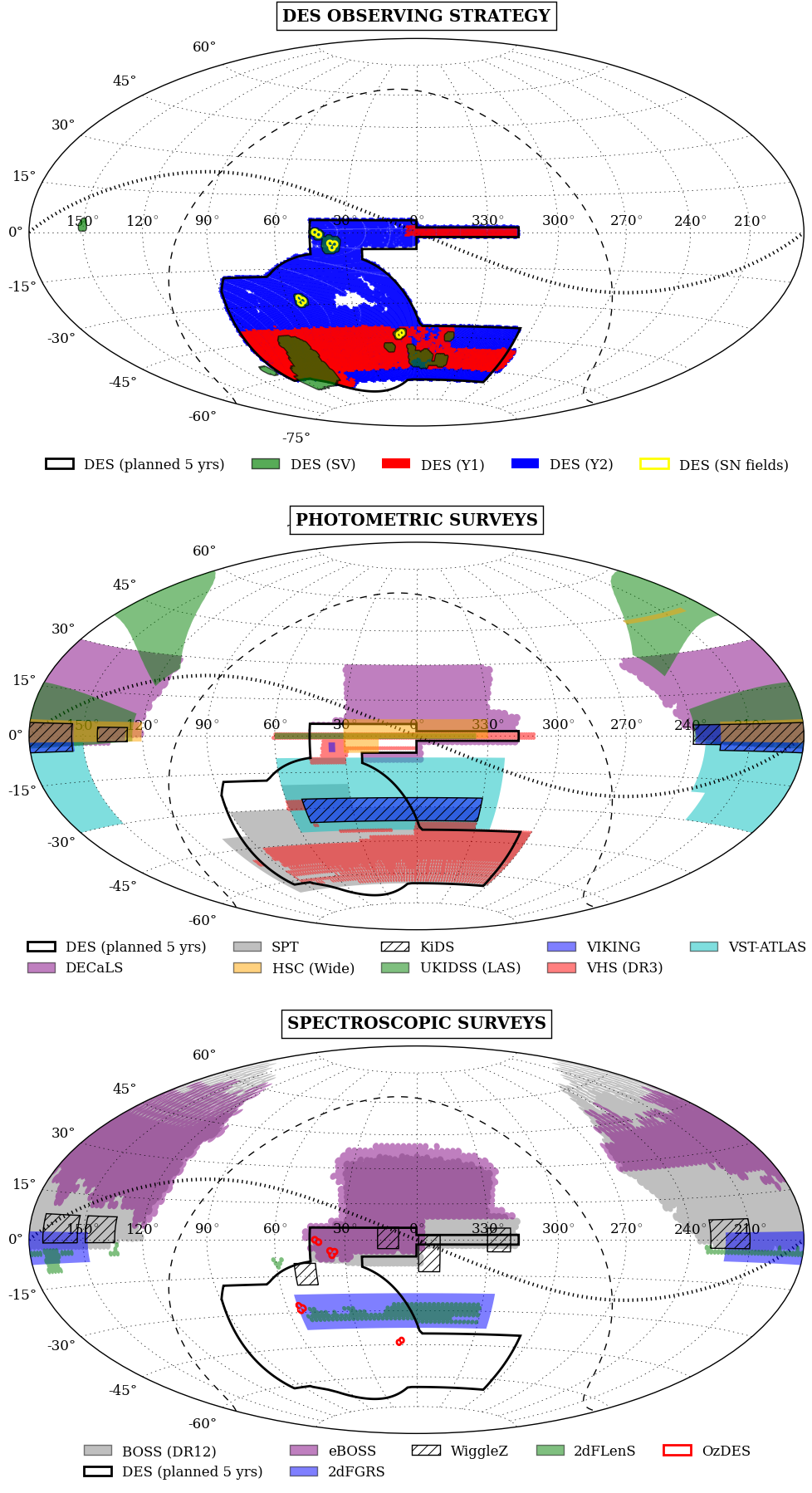}
 \caption{DES and some selected completed or ongoing surveys (as of December 2015). 
This is a Hammer projection in equatorial coordinates, with the dashed and dotted lines indicating the Galactic plane and the ecliptic plane, respectively. 
{\it Top}: DES survey footprint  for the SV, Y1, Y2 and the final 5-yrs survey; 
{\it Middle}: with other  photometric surveys; 
{\it Bottom}: with other spectroscopic surveys. 
Planned wide-field surveys not shown include LSST and WFIRST (both photometric),  Euclid  (photometric and spectroscopic), Sumire PFS, DESI and 4MOST (all three spectroscopic).
We note the VHS (DR3) indicates deep coverage in the DES area. Eventually it will cover nearly the whole of the DES 5-yrs footprint.
A shallower VISTA whole-hemisphere is also underway. The DES footprint was designed to have large overlap with SPT (shown here in the 'photometric surveys' plot).
DES SV and Y1 areas essentially completely overlap SPT, and the full planned DES overlaps the completed 2500 sq deg SPT.
}
\label{fig:footprints}
\end{figure*}

\begin{table*}
\caption{The current DES inventory: different objects observed with DES over the SV, Y1 and Y2 seasons, and the expectation for 
the full 5 seasons.  $\sigma$ for galaxies and stars indicates signal/noise ratio, $\lambda$ for clusters denotes richness.}
\begin{tabular}{@{}ccc}
\toprule
Objects & As of December 2015 & Expected from full 5yr DES \\
\toprule
Galaxies with photo-\emph{z} ($ > 10 \sigma$) & 7M (SV), 100M (Y1+Y2) & 300M\\
Galaxies with shapes & 3M (SV), 80M (Y1+Y2) & 200M\\
Galaxy clusters ($\lambda > 5$) & 150K (Y1+Y2) & 380K\\
\midrule
SN Ia & 1000 & Thousands  identified photometrically\\
Super-luminous SN & 2 (published)  + more  confirmed spectroscopically + many more candidates & 15-20 \\
\midrule
New Milky Way companions & 17 & 25\\
\midrule
QSOs at $z>6$ & 1 (published)   + more  confirmed spectroscopically  + many more candidates  & 375\\
Lensed QSOs & 2  & 100 ($i<21$)\\
\midrule
Stars ($ >  10 \sigma$) & 2M (SV), 30M (Y1+Y2) & 100M\\
\midrule
Solar System: new Trans Neptunian Objects & 32 in SN fields + 2 in the WF & 50 + many more in the wide field\\
New Jupiter Trojans & 19 & \\
Main Belt Asteroids & 300K (Y1+Y2) & \\
Kuiper Belt Objects & & 500-1000\\
\toprule
\end{tabular}
\label{tab:DES inventory}
\end{table*}

DES is an imaging survey of 5000 sq deg of the Southern sky, using a 570 megapixel camera on the Cerro Tololo Inter-American Observatory (CTIO) 4m Blanco telescope in Chile. Photometric redshifts are obtained from the multi-band photometry to produce a quasi-three dimensional survey. The main goal of DES is to determine the DE equation of state $w$ (the ratio of pressure to density) and other key cosmological parameters to high precision. DES will measure $w$ using  complementary techniques in a single survey: counts of galaxy clusters, 
weak gravitational lensing, and galaxy power spectra. DES is expected to catalogue 300 million galaxies with photometric redshifts; 200 million of them will have  shape measurements for weak lensing.  In addition, 
thousands of  Type Ia supernovae  will be measured in 27 sq deg `time domain' fields. It is expected that the uncertainty on $w$ will be only a few percent for each probe. See \citet{DES2005} and a companion overview paper (DES Collaboration, in preparation) for detailed parameterizations and statistics.

DES is an international collaboration, with over 400 scientists from  the US, the UK, Spain, Brazil, Germany,  Switzerland and Australia.
The DES science is coordinated by a Science Committee comprised of thirteen Science Working Groups (SWGs).
Core DE SWGs include large scale structure, clusters, weak lensing and supernovae Ia.
Additional SWGs which focus on the primary science are photometric redshifts, spectroscopy, simulations, and  theory \&  combined probes. 
The Non-DE SWGs focus on Milky Way science, galaxy evolution, strong lensing, quasars, and transients \& moving objects.

The DES footprint is shown in Figure~\ref{fig:footprints}, and Table~\ref{tab:DES inventory} presents an inventory of measured and discovered objects so far (December 2015), as well as forecast for the complete survey. 
The first light of DES was obtained in September 2012,
followed by  commissioning of DECam \citep{Flaugher2015}.
SV observations took place from November 2012 to February 2013, and the Y1 and Y2  observations were taken from August 2013 to February 2014, and from August 2014 to February 2015, respectively. 
The SV data cover  about 250 sq deg to nearly the full DES depth, in part overlapping the South Pole Telescope (SPT) 
area, as well as some supernovae fields. Analyese presented below have used different subsets of the SV data. 
Five optical filters ($grizY$) were used throughout the
SV, with typical single exposure times of 90 seconds for $griz$
and 45 seconds for $Y$. 
Details of the resulting magnitude limits from the SV season 
in the wide survey in each band (nominally for 10  exposures, but in practice fewer) are shown in Table~\ref{tab:exposures}. 
For the depth obtained during SV with the supernovae fields see \citet{des_cs_2014}.

Separation of stars and galaxies in the SV data was done using a number of codes, resulting in a galaxy purity of 98\% and completness of 
80\% down to $i=22.5$ \citep{Crocce2015}. See more in \citet{soumagnac2015}  for star/galaxy separation methodology for DES.
The performance of several photo-\emph{z} methods applied to SV data was evaluated in \citet{des_cs_2014}.
The best methods yielded scatter $\sigma_{68}  = 0.08$  (defined as the 68\%  width about the median of  $\Delta z  = z_{\rm spec}  - z_{\rm phot}$). 
Regarding the image quality, the achieved median seeing FWHM is about 0.9'' in filters {\it  riz}, as  expected when designing the survey for weak lensing analyses. 
 The shape measurements (using two methods) for about 3 million  SV  galaxies are presented in 
\citet{des_mj_2015}, who conclude that the shear is determined with multiplicative errors of $\pm5\%$ (1-sigma). 
The 27 sq deg (over 10 fields) that comprise the supernova
survey \citep{Bernstein2012} are observed in the $griz$ bands at approximately weekly intervals throughout each six-month DES observing season, and provide a much more densely sampled
time series.


\begin{table}
\begin{center}


 \caption{
Estimated PSF magnitude limits from a subset of the  DES Science Verification (SV) data, based on Balrog simulations embedded into 210 sq deg of the DES 
imaging \citep[see][]{des_es_2015}.
We calculate the S/N vs. magnitude response in tiles of approximately 0.5 sq deg, identify the S/N=10 magnitude value in each tile, then compute the median of these values. The estimation is for objects simulated as point sources (regardless of their measured classification as stars or galaxies). The DES SV survey 
mimicked the full 5-year DES survey (ideally of 10 exposures), but with significant depth variations due to weather and other challenges during early operations of DECam \citep[see e.g.][]{Leistedt2015}, so effectively less than 10 exposures.  
Overall the Blanco and DECam performance has improved considerably after  SV.
\label{tab:exposures}
}
 \begin{tabular}{@{}cccccc}
 \hline
 band & Nominal SV exposures &  mag (10$\sigma$)   \\ \hline
 $g$  & $10 \times 90$ sec& 24.45  \\
 $r$   & $10 \times 90$ sec& 24.30  \\ 
 $i$  & $10 \times 90$ sec & 23.50  \\
 $z$  & $10 \times 90$ sec & 22.90  \\
 $Y$ & $10 \times 45$ sec & 21.70\\ \hline
 \end{tabular}
\end{center}
\end{table}


Although observed for testing purposes, SV is a powerful dataset
in its own right, particularly for weak lensing, where it rivals the
full CFHTLenS survey for size (though DES SV is slightly
shallower). By design, there is significant overlap with the SPT survey (see Figure~\ref{fig:footprints} for overlapping surveys).
The SV phase of DES was recently combined with near infrared data from ESO's VISTA Hemisphere Survey (VHS) ,
described in \citet{McMahon2013} and \citet{Banerji2015}, which was used to extract fluxes and associated errors from the shallower VHS data. Joint 7-band ({\it grizYJK}) photometric catalogues were produced for a single 3 sq deg DECam field, where the multi-wavelength photometry and spectroscopy were used for data quality tests. This will be extended to the full DES+VHS when the survey is complete. 
We also note the synergy with the Australian OzDES, a 100-night spectroscopic survey with the 2dF fibre-fed spectrometer on the Anglo-Australian Telescope dedicated to following-up DES targets identified in the SN fields \citep{Yuan2015}.

The Y1 and Y2 observations (now complete) aimed at covering most of the 5000 sq deg footprint to about 40\% of the final survey depth. Single-epoch data of Y1\&Y2 have already been used in searches  for new Milky Way companions \citep{Bechtol2015, des_adw_2015}.
Catalogues are currently being prepared for studies of large scale structure, clusters and weak lensing. As shown in Table 1,  the Y1 \&Y2 catalogues will yield 100M galaxies with photo-\emph{z}, of which 
80M
will have shape measurements for weak lensing studies. 

The collaboration also prepared a set of Data Challenge simulations (Data Challenge 6, BCC, MICE), which are being analysed alongside the real data as explained below. The simulations are described in other collaboration papers.


\section{Solar System Science with \textsc{DES}}



\label{sec:SSmain}

DES was designed to have a combination of area and depth optimized for extragalactic astronomy and cosmology. To achieve the deep co-added images
required for cosmological measurements, the same fields are repeatedly observed over the five years of the survey. This survey strategy also 
gives DES the ability to detect moving transients over $\sim1/8$ of the sky, making it a uniquely powerful tool for observation and discovery within the solar system. 

The ability to detect solar system transients is a function of both the single-epoch exposure depth and the observing cadence, as well as the rate of motion of the object through
the survey area. 
Most solar system objects follow orbits within $\pm 20^{\circ}$ of the ecliptic plane. Higher-inclination objects have typically experienced at least one 
scattering event, and their dynamics encodes the history of such encounters. Figure~\ref{fig:footprints} shows the wide survey and supernova
fields in relation to the ecliptic plane. The equatorial stripe encompasses approximately 500 sq deg within 20 degrees of the ecliptic. Five of the supernova fields
also lie at moderate ecliptic latitudes. The bulk of the survey area, however, lies at ecliptic latitudes below $-20^{\circ}$. This region is seldom covered by surveys that explicitly
target solar system objects---for example the `other' DES, the Deep Ecliptic Survey \citep{DES-1}---but makes \textit{this} DES particularly sensitive to unusual high-inclination objects, such as Centaurs and members of the distant scattered disk.

We illustrate the capabilities of DES with a discussion of four types of solar system objects: near-earth objects, main belt asteroids, distant objects including Trojans and Centaurs, and 
the trans-Neptunian region. 

\subsection{Near-Earth Objects}
\label{sec:NEO}
Near-earth objects (NEOs) move rapidly and are often observed over arc lengths of only a few days. For this reason, their positional uncertainty can grow rapidly, and many newly-detected NEOs
soon become undetectably faint as they recede from Earth and hence are lost. DECam's depth and 3 sq deg field of view make it an excellent instrument for NEO detection and recovery. DES has performed 
target-of-opportunity observations of a handful of NEOs, such as the recovery of the potentially hazardous asteroid 2014~BE$_{63}$.
A programme to search for NEOs with DECam is underway using community time \citep{DECam-NEO}.

\subsection{Main-Belt Asteroids}
\label{sec:MainBelt}
The vast majority of known asteroids occupy the main asteroid belt between Mars and Jupiter. As of 6 Jan 2015, the Minor Planet Center\footnote{\url{http:/www.minorplanetcenter.net}} (MPC) lists over 650,000 objects with semi-major axes between 2 and 4~AU. By comparing the calculated positions and magnitudes of these asteroids with the $\approx$40,000 DES wide and supernova survey exposures from
 the Y1 and Y2  observing seasons,
we estimate that approximately 70,000 known main belt asteroids should already have been observed by DES, with detected objects being visible in a median of six exposures at a median
magnitude of 20.5. Hence DES can already provide useful information about a large number of main belt asteroids.

\begin{figure*}
\begin{minipage}{145mm}
\includegraphics[width=\columnwidth]{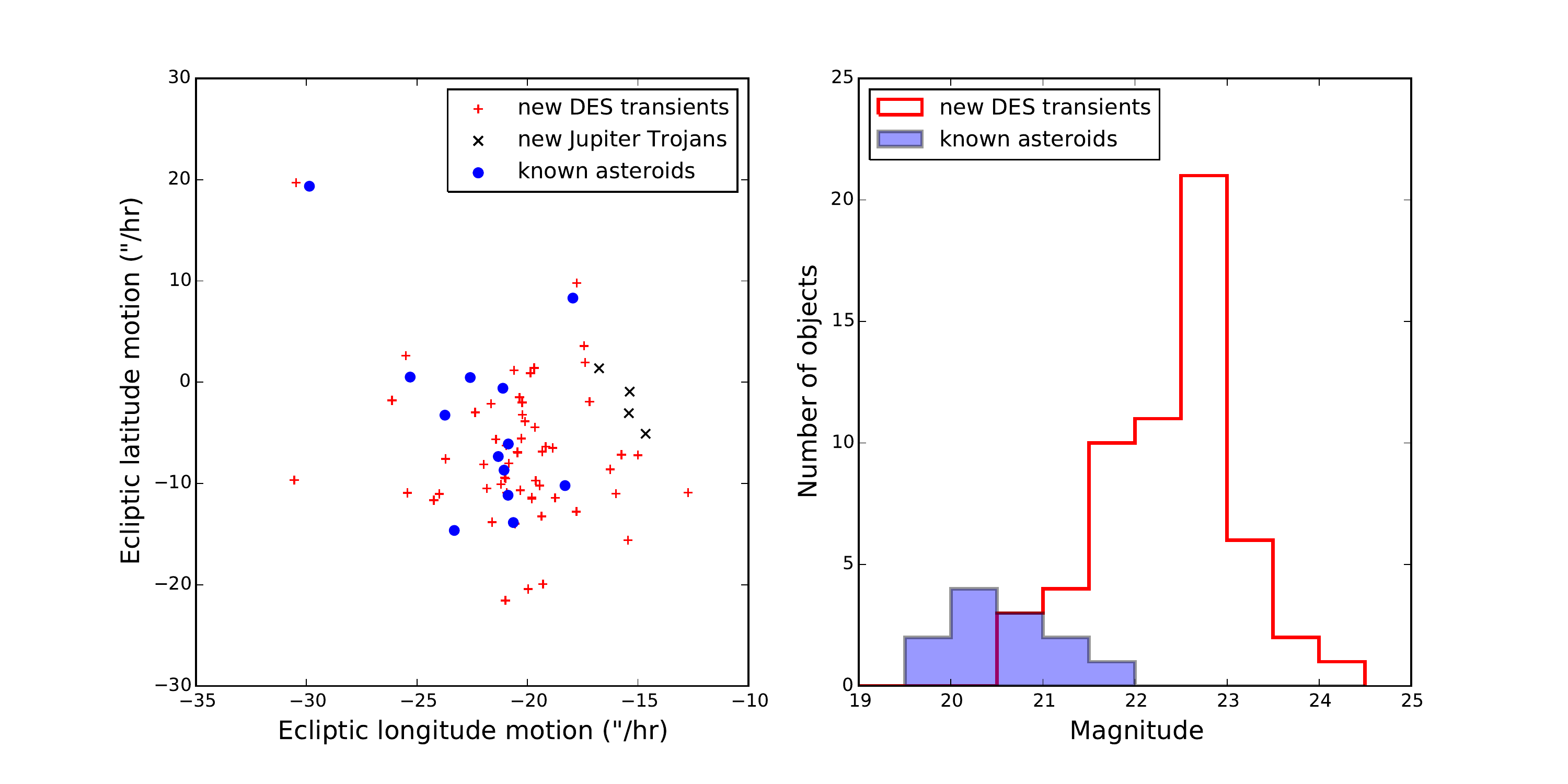}
 \caption{Properties of transients observed in a single visit to the DES supernova search field X3, consisting of five 360-second \emph{i}-band exposures on 24 Sept 2013. Left: hourly motion in ecliptic latitude and longitude, with
 objects matched to previously-known main belt asteroids indicated. The new DES detections include several Jupiter Trojans that are among 19 such objects found by DES to date. Right: Comparison of magnitudes of previously-known objects with new DES detections. The known asteroids have a median \emph{i}-band magnitude of 20.5, compared to a median of 22.6 for the new DES objects.}
  \label{fig:hourlymotion}
 \end{minipage}
\end{figure*}

However the catalogues of known solar system objects are very incomplete at the DES exposure depth. To illustrate this, the left panel of
Figure~\ref{fig:hourlymotion} shows the motion, in arc seconds per
hour, of moving transients detected in a single exposure sequence in the deep supernova field X3 on 24 September 2013. This sequence consisted of five consecutive 360-second \emph{i}-band  exposures. Seventy-three
moving transients were detected in this sequence, of which only 13 were matched to known main belt asteroids. The rate of retrograde motion of the unmatched population is consistent with the known objects, suggesting that they are also main belt asteroids. There is, however, an excess of unmatched objects with somewhat lower motions. Several of the latter objects have been observed in images taken during other epochs and confirmed as Trojan companions of Jupiter.  

Though still at an early stage, the search for main belt asteroids with DES should eventually lead to hundreds of thousands of new detections. Linking single-night observations of new main belt objects into orbits
poses a challenge, however, because the DES wide-survey observing cadence is not well-matched to these objects' relatively high rates of motion. There is currently no plan within DES to attempt orbit-linking for main belt asteroids, but we intend to submit all detected moving transients to the Minor Planet Center, where links may be made to current or future data from other surveys.

\subsection{Distant Solar System Objects}
\label{sec:distant}

The solar system beyond the main belt contains several distinct groups of minor planets:
\begin{itemize}
  \item The \textbf{Trojan populations} of Jupiter and Neptune, consisting of objects in a stable 1:1 mean-motion resonance centred upon the  leading (L4) or trailing
(L5) Lagrange points of their parent body.
\item \textbf{Centaurs}, with chaotic orbits that cross the orbits of the giant planets. Believed to be 
objects ejected from the Kuiper Belt, Centaurs evolve along complicated dynamical pathways that may end in capture as a Jupiter-family comet, collision with a giant planet or with the Sun, or even ejection 
from the solar system. 
\item \textbf{The trans-Neptunian region}, with a rich dynamical structure that includes the classical Kuiper Belt, various mean-motion Neptunian resonances, the scattered disk, and a small but very interesting group of detached / inner Oort cloud objects with perihelia greater than 50~AU (such as the dwarf planet Sedna).  An early indication of the power of DECam came with the discovery by \citet{Trujillo2014} 
    of a second Sedna-like object, 2012~VP$_{113}$, whose perihelion of 80.3~AU is the largest known. 
\end{itemize} 

\begin{figure*}
\includegraphics[width=\textwidth]{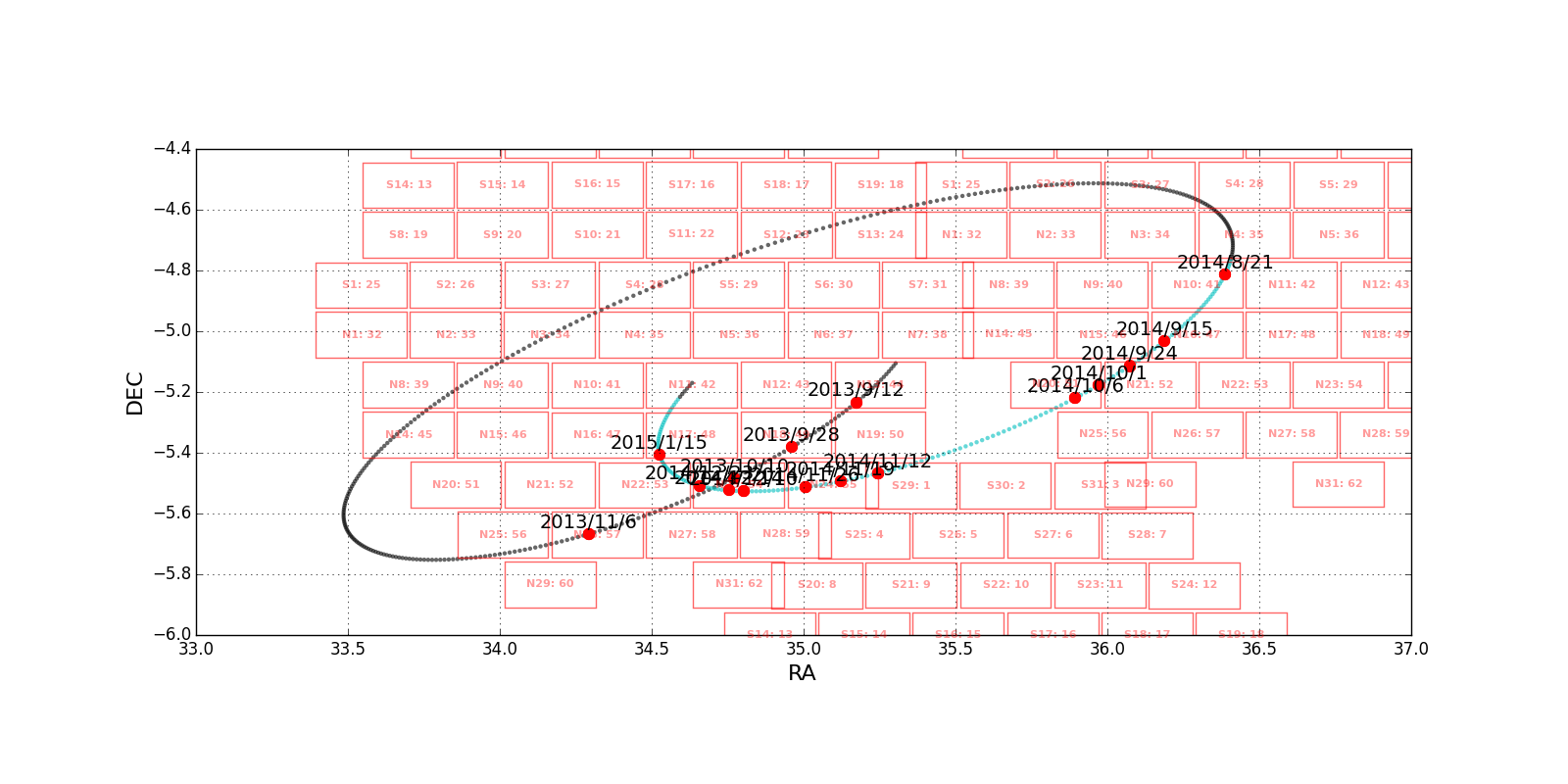}
 \caption{Trajectory of the plutino 2013 RD$_{98}$ through three adjacent supernova fields during the DES Y1 and Y2 seasons. Large dots indicate nights on which
 the object was observed by DES. Each rectangle represents one DECam CCD.}
  \label{fig:TNO-linking}
\end{figure*}

To date the search for TNOs with DES has been carried out in the ten supernova fields using the full Y1 and Y2 data samples, as well as the Y3 data up through October 2015.
These fields encompass 27 sq deg, less than 1\% of the full wide-survey area, but are an excellent starting point for the TNO search because they are observed with a one week nominal cadence in the $griz$ bands throughout each six-month DES observing season. 
 Five of the fields are at moderate ecliptic latitudes of $-15$ to $-20$ degrees, while the other five are at higher latitudes of around $-45$~degrees. The search begins with the list of transient candidates identified 
in each single-epoch exposure using the supernova difference-imaging pipeline \citep{Kessler2015}. Artifacts and other non-point-like detections are removed with the Autoscan machine-learning algorithm of \citet{Goldstein2015}. The apparent motion of a TNO is dominated by earth-reflex motion; for example an object at 40~AU observed near opposition will undergo retrograde motion at 3-4$^{\prime\prime}$/hour. This allows us to identify candidate triplets of points---the minimum number required for an orbital solution---from nearby visits to a given field, separated in a way consistent with seasonally appropriate reflex motion. We attempt to fit each triplet to an orbit using the \texttt{fit\_radec} algorithm of \citet{Bernstein2000}. With a 
successful candidate orbit in hand, the two-week arc of a typical triplet can be readily extended to include points from subsequent or prior visits, into adjacent fields including possible wide-survey detections, and into 
different DES observing seasons. The trajectory of one such object is shown in Figure~\ref{fig:TNO-linking}.

\begin{figure}
\includegraphics[width=\columnwidth]{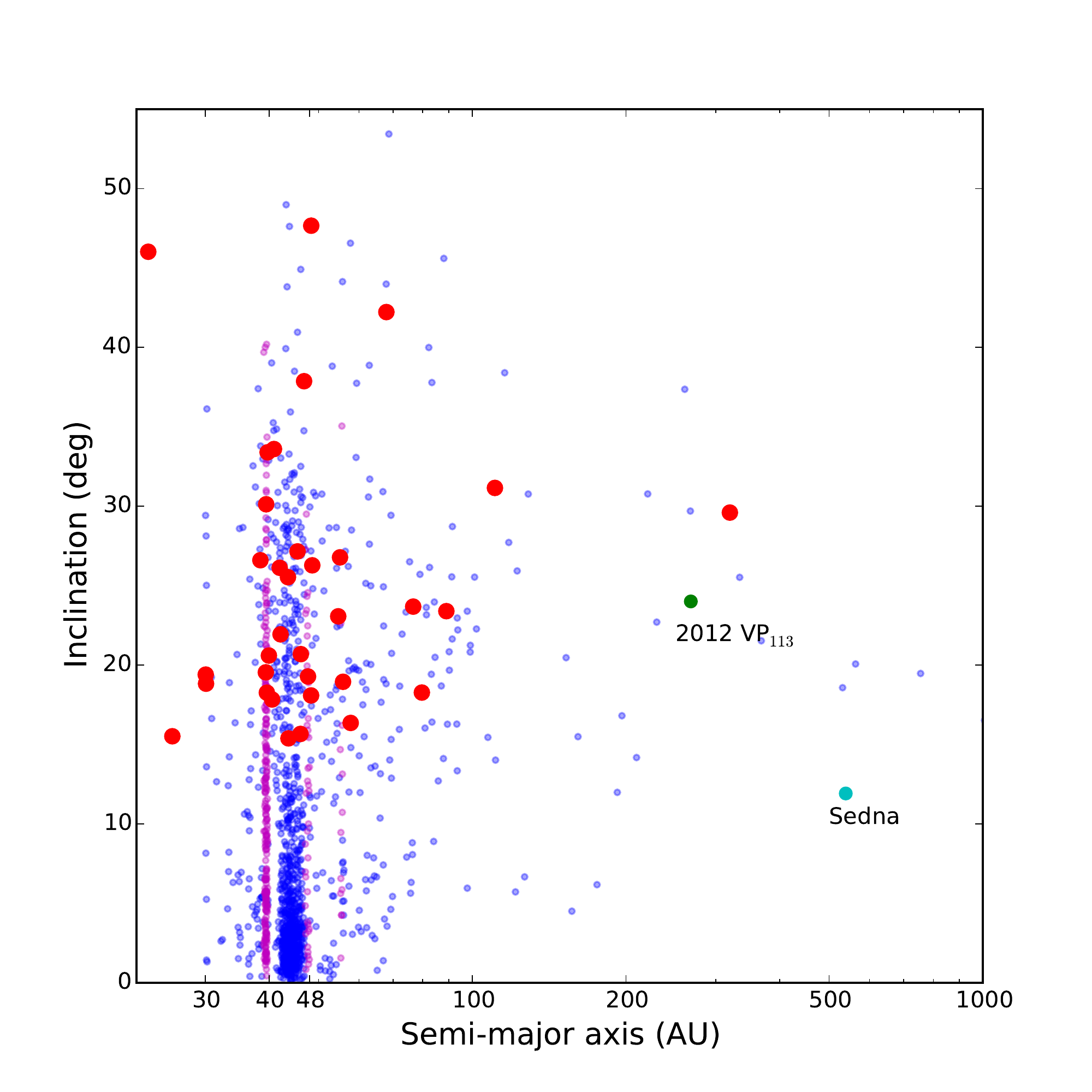}
\caption{Inclination vs. semi-major axis for 32 distant solar system objects (large circles) discovered in the DES supernova fields. Small circles indicate other 
 known TNOs. The band of objects at $a=39.3$~AU are plutinos in a 3:2 mean motion resonance with Neptune. The inner Oort cloud objects Sedna and 2012~VP$_{113}$ are also indicated. Due to the location of the search fields, all of the new detections have inclinations of at least 15$^{\circ}$. }
\label{fig:TNOs-inclination-vs-a}
\end{figure}

\begin{figure}
\includegraphics[width=\columnwidth]{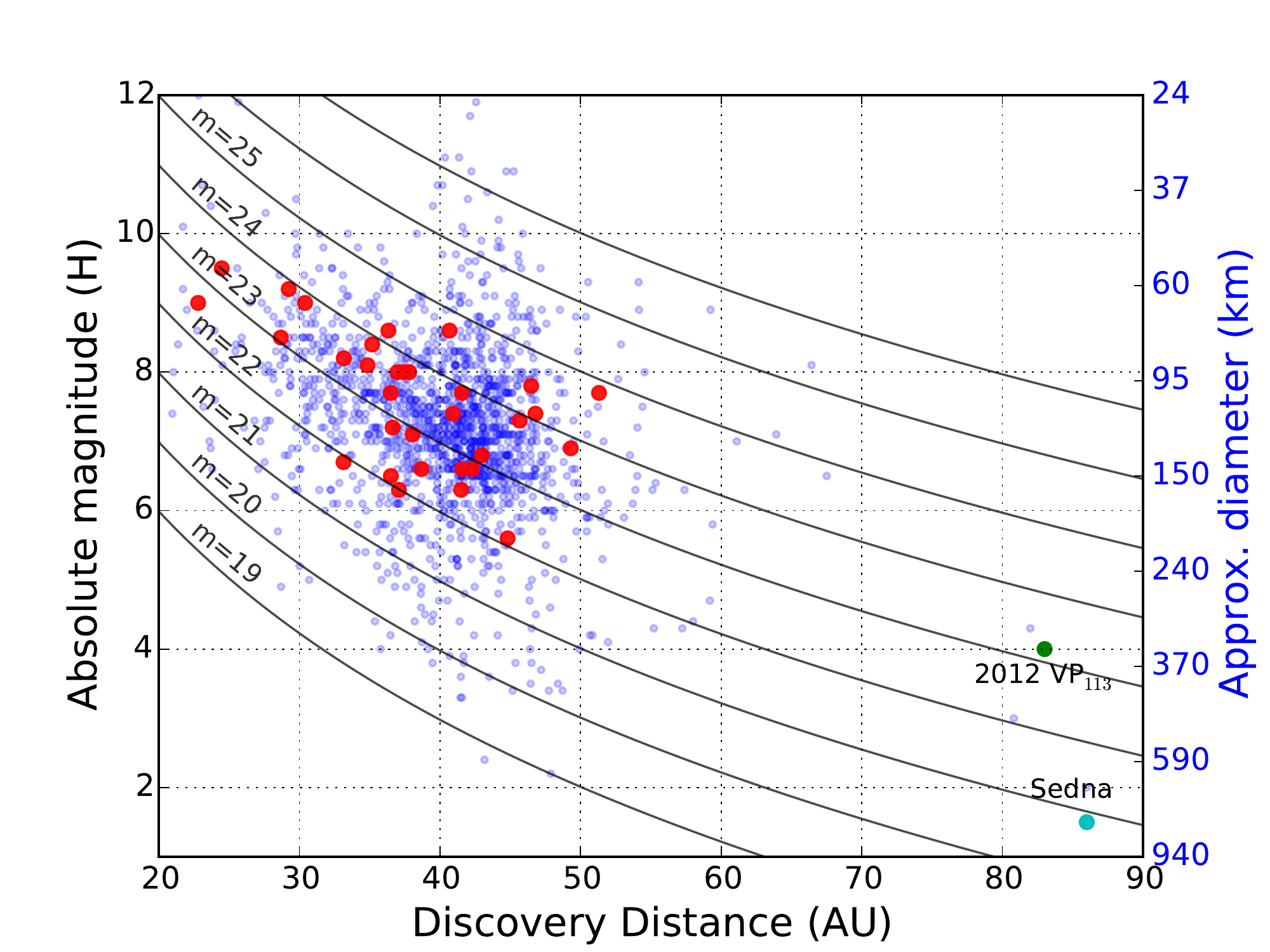}
 \caption{Absolute magnitude vs. heliocentric distance at discovery for 32 distant solar system objects (large circles) discovered in the DES supernova fields. Small circles indicate other 
 known TNOs. The conversion from absolute magnitude to size assumes an albedo of 5\%. The inner Oort cloud objects Sedna and 2012~VP$_{113}$ are also indicated. }
  \label{fig:TNOs-mag-vs-dist}
\end{figure}

The search for TNOs in the DES supernova fields has yielded 32 new objects through the first part of Y3. 
Their orbital elements and other properties  are listed in Table~\ref{tab:TNOs}. Due to the location of our search fields, all have inclinations
of at least $15^{\circ}$, increasing the population of known ``hot'' TNOs by about 10\%.  Figure~\ref{fig:TNOs-inclination-vs-a} shows the inclination vs. semi-major axis for these new objects, while
Figure~\ref{fig:TNOs-mag-vs-dist} shows absolute magnitude as a function of solar distance at discovery. From this figure, one can read off the sensitivity of DES to 
larger, more distant objects; e.g. with a magnitude depth of 24, DES could discover a $H=5$ object (approximately 600~km in diameter) at  80~AU, a distance comparable to the distance at discovery
of the larger objects Sedna and 2012~VP$_{113}$.  The DES discoveries include a number of objects in mean motion resonances
with Neptune, as well as several distant scattered disk objects. In particular we highlight the pair of L4 Neptune Trojans 2014~QO$_{441}$ and 2014~QP$_{441}$ \citep{Gerdes2015}, which raise
the number of known stable Neptune Trojans to twelve. We also highlight
the ``extreme TNO'' 2013~RF$_{98}$, whose 325~AU semi-major axis is the fifth-largest of any known solar system object that does not cross the orbit of Neptune.
2013~RF$_{98}$ and the other longest-period TNOs display as-yet-unexplained clustering in the argument of perihelion near $\omega=0^{\circ}$, which may result from a stabilizing interaction
with a distant perturber \citep{Trujillo2014, FuenteMarcos2014}.
The minor planet 2013~RF$_{98}$ is one of the six most distant TNOs whose aligned orbits were used by 
\citet{BatyginBrown2016}
to hypothesize the existence of Planet 9.
As the modelled trajectory of Plant 9 lies within the DES footprint, DECam would be ideal  to search for it.

While the TNO search in the supernova fields has already produced objects of interest and will continue to do so in the remaining years of the survey, it is ultimately the DES wide survey
that will have the greatest potential to chart new territory in the trans-Neptunian region.
With its combination of area, depth, and a survey footprint weighted toward high ecliptic latitudes, DES has the potential to discover $\sim30\times$ more
``hot'' TNOs than any previous survey. DES's high-latitude coverage compliments Sheppard and Trujillo's ongoing TNO search programme using DECam at lower ecliptic latitudes.
Although little is known about the distribution of detached Sedna-like objects, it seems reasonable to suppose that they are less likely to lie
 close to the ecliptic plane than more typical TNOs. If so, DES has the potential to discover tens of such objects, making detailed studies of the population and dynamics of the inner Oort
 cloud possible for the first time.

\begin{table*}
\begin{center}
 \caption{Barycentric osculating orbital elements and other properties of 32 distant solar system objects discovered in the DES supernova fields. These data were
 retrieved from the Minor Planet Center on 20 Nov 2015. Scattered disk objects (SDOs) and objects in mean motion resonances with Neptune are indicated in the last column.}
 \label{tab:TNOs}
 \begin{tabular}{@{}lccccccccccc}
     \hline
     Designation & $a$ (AU) & $e$ & $i$ (deg.) & $\Omega$ (deg.) & $\omega$ (deg.) & $M$ (deg.) & epoch & peri. date & $H$ & comment \\ \hline
2012 VR$_{113}$ & 47.57 & 0.17 & 19.3 & 121.1 & 220.5 & 38.2 & 2015-06-27.0 & 1980-09-08 & 6.6 & 2:1 resonance \\
2012 VS$_{113}$ & 55.02 & 0.31 & 26.8 & 171.7 & 220.4 & 3.9 & 2015-06-27.0 & 2011-02-09 & 7.1 & SDO \\
2012 VU$_{113}$ & 39.43 & 0.08 & 30.1 & 65.5 & 315.8 & 12.5 & 2015-06-27.0 & 2006-11-25 & 7.7 & plutino \\
2012 VV$_{113}$ & 46.01 & 0.11 & 15.7 & 116.1 & 295.6 & 349.3 & 2015-06-27.0 & 2024-10-08 & 7.4 & \\
2012 WD$_{36}$ & 76.52 & 0.51 & 23.7 & 177.4 & 293.9 & 336.8 & 2016-01-13.0 & 2059-03-02 & 6.9 & \\
2012 YO$_{9}$ & 43.64 & 0.17 & 15.4 & 118.7 & 3.3 & 298.7 & 2015-06-27.0 & 2064-07-16 & 7.7 & SDO \\
2013 QO$_{95}$ & 39.89 & 0.03 & 20.6 & 83.1 & 323.2 & 345.4 & 2015-06-27.0 & 2025-09-10 & 6.6 & \\
2013 QP$_{95}$ & 40.57 & 0.17 & 25.4 & 71.4 & 19.5 & 314.5 & 2015-06-27.0 & 2048-03-01 & 7.2 & 11:7 resonance \\
2013 RB$_{98}$ & 42.17 & 0.26 & 22.0 & 147.2 & 350.6 & 286.5 & 2015-06-27.0 & 2071-06-02 & 6.6 & 5:3 resonance \\
2013 RD$_{98}$ & 39.36 & 0.23 & 19.6 & 145.0 & 104.0 & 126.2 & 2015-06-27.0 & 1928-11-14 & 7.3 & plutino \\
2013 RF$_{98}$ & 325.1 & 0.89 & 29.6 & 67.6 & 316.5 & 0.1 & 2013-09-25.0 & 2011-11-18 & 8.6 & SDO \\
2013 RG$_{98}$ & 23.22 & 0.17 & 46.0 & 146.0 & 350.3 & 291.6 & 2015-06-27.0 & 2036-10-01 & 9.0 & Centaur \\
2013 SE$_{99}$ & 79.54 & 0.54 & 18.3 & 100.6 & 223.8 & 20.8 & 2015-06-27.0 & 1974-07-15 & 7.8 & SDO \\
2013 TH$_{159}$ & 40.86 & 0.09 & 33.6 & 62.4 & 233.0 & 84.3 & 2015-06-27.0 & 1954-05-01 & 8.6 & \\
2013 TV$_{158}$ & 111.24 & 0.67 & 31.2 & 181.1 & 232.3 & 357.8 & 2015-06-27.0 & 2022-08-22 & 6.3 & SDO \\
2013 VD$_{24}$ & 54.51 & 0.24 & 18.9 & 105.7 & 196.8 & 62.6 & 2013-12-14.0 & 1943-12-16 & 7.7 & SDO \\
2014 QL$_{441}$ & 48.36 & 0.25 & 26.3 & 75.8 & 293.5 & 11.6 & 2016-01-13.0 & 2005-03-25 & 6.5 & \\
2014 QM$_{441}$ & 39.5 & 0.28 & 18.3 & 96.1 & 313.0 & 354.9 & 2016-01-13.0 & 2019-07-16 & 8.5 & plutino \\
2014 QN$_{441}$ & 46.07 & 0.28 & 20.7 & 144.7 & 242.8 & 3.1 & 2015-06-27.0 & 2012-09-30 & 6.7 & \\
2014 QO$_{441}$ & 30.06 & 0.11 & 18.8 & 107.1 & 112.5 & 169.9 & 2015-06-27.0 & 1937-09-12 & 8.2 & Neptune Trojan \\
2014 QP$_{441}$ & 30.08 & 0.07 & 19.4 & 96.6 & 3.2 & 297.9 & 2014-10-30.0 & 2043-04-15 & 9.2 & Neptune Trojan \\
2014 QR$_{441}$ & 67.45 & 0.37 & 42.2 & 77.9 & 283.0 & 354.6 & 2016-01-13.0 & 2024-04-18 & 6.8 & SDO \\
2014 QS$_{441}$ & 46.76 & 0.06 & 38.0 & 186.0 & 257.2 & 316.6 & 2016-01-13.0 & 2054-07-19 & 5.6 & \\
2014 QU$_{441}$ & 41.15 & 0.11 & 19.3 & 113.2 & 301.2 & 341.0 & 2016-01-13.0 & 2029-12-18 & 8.0 & \\
2014 SB$_{349}$ & 25.82 & 0.18 & 15.5 & 117.2 & 359.2 & 300.4 & 2016-01-13.0 & 2037-10-01 & 9.5 & Centaur \\
2014 SZ$_{348}$ & 48.07 & 0.23 & 47.7 & 158.6 & 239.6 & 6.9 & 2015-06-27.0 & 2009-02-12 & 8.0 & \\
2014 TT$_{85}$ & 40.58 & 0.07 & 17.8 & 98.4 & 280.8 & 16.1 & 2014-10-30.0 & 2003-03-29 & 8.0 & \\
2014 TU$_{85}$ & 48.65 & 0.31 & 16.4 & 120.8 & 306.1 & 340.9 & 2014-10-10.0 & 2032-10-21 & 8.4 & \\
2014 UF$_{224}$ & 45.39 & 0.13 & 27.2 & 174.4 & 330.5 & 264.7 & 2015-06-27.0 & 2096-06-08 & 7.4 & \\
2014 VT$_{37}$ & 39.95 & 0.26 & 33.5 & 197.7 & 175.6 & 18.6 & 2014-12-09.0 & 2001-11-25 & 9.0 & plutino \\
2015 PD$_{312}$ & 54.54 & 0.36 & 23.1 & 155.0 & 243.7 & 356.9 & 2015-08-26.0 & 2019-01-26 & 8.1 & SDO \\
2015 PF$_{312}$ & 45.39 & 0.1 & 18.0 & 160.9 & 208.3 & 24.1 & 2015-08-26.0 & 1995-02-22 & 6.3 & \\ \hline
 \end{tabular}
\end{center}
\end{table*}

\section{Stellar, Milky Way, and Local Group Science}
\label{sec:intro}


Wide-field photometric surveys have significantly expanded our understanding of the Milky Way and its stellar populations. For instance, the speed and depth of recent digital surveys have considerably improved the census of low-mass and low-luminosity stars and sub-stellar objects~\citep[\eg,][]{Zhang2009, Kilic2006}. 
These surveys have also revealed complex substructure in the Galactic halo~\citep[\eg,][]{RochaPinto2004, Willman2005, Koposov2008, Walsh2009, Sharma2010}.
Together, these data sets have allowed us to better characterize the structure, mass assembly, and star formation history of our Galaxy~\citep[\eg,][]{Juric2008, Reyle2009}. 
With exceptional sensitivity over the southern Galactic cap, DES promises to further expand our knowledge of the local environment.

The purpose of this section is to address topics in stellar, Milky Way, and Local Group science where the DES survey will provide important input. 
We will address topics in order of proximity to the Earth, starting with nearby Milky Way stellar populations and concluding with Local Group satellite galaxies.
When possible, we will note topics that are already being studied with the early DES data sets.
The list of topics cited here is almost certainly incomplete, and will likely be expanded as the full DES data set becomes available.
Much of the work in this section builds on the studies of \citet{Rossetto2011}; however, the DES footprint and projected photometric sensitivity have been subsequently refined.

\subsection{A census of L and T dwarfs in the Galactic disk}
\label{sec:lt_dwarfs}
 A large number of substellar objects are necessary to infer the fundamental properties of brown-dwarf populations, such as their initial mass function (IMF). Unlike normal main-sequence stars, determining the masses of brown dwarfs requires both their luminosity and ages to be known, effectively coupling the determination of their IMF and Star Formation History (SFH) \citep{DayJones2013b}. The \Trilegal Galactic model adopted by~\citet{Rossetto2011} predicts that DES will observe $\roughly 3 \times 10^4$ objects with $M \leq 0.08 \Msolar$. 
This number is about an order of magnitude larger than the current census of known objects of sub-stellar mass.
Given the extremely low luminosities of these objects, they will be largely restricted to within one or two thin disk scale heights. 
It will be challenging to extract the sample of L and T dwarfs from the DES data, partly due to these objects being detectable only in the redder filters. 
Availability of the \band{\emph{Y}}-band will be helpful in disentangling them from the high-\emph{z} galaxy and QSO populations, specifically based on their positions in \colourp{\emph{z}}{\emph{Y}} \vs \colourp{\emph{i}}{\emph{z}} colour-colour diagrams. 
DES and VHS data can be combined to provide a more comprehensive fit to the spectral energy distributions of these objects.  

A complementary approach to finding nearby substellar objects is to use proper motion measurements. 
A local disk star with a transverse velocity of $V_T = 20\km\second^{-1}$ relative to the local standard of rest and a distance of $d = 50\pc$ has a proper motion of $\mu \simeq 80 \mas \yr^{-1}$. 
Given the 5-year baseline of the DES survey, the resulting displacement of $0.4$ arcsec is expected to be measurable with the DES data.
DES obtains about 15 mas relative astrometric accuracy per exposure for bright objects, limited by atmospheric distortions.

The large number of L and T dwarf candidates will provide excellent targets for spectroscopic follow up, which may help to constrain atmospheric models of these objects. 
We note that the \citet{Chabrier2000} models used in the simulations by \citet{Rossetto2011} typically have colours which are systematically blue compared to those observed in (Sloan Digital Sky Survey) SDSS \citep{Knapp2004,DayJones2013a}, even after differences in transmission between DES and SDSS filters are taken into account. 
More recent models from \citet{Saumon2008} display redder colours by $\simeq 0.2$ mag, but still have trouble reproducing the observed colour locus for later-type objects.

\subsection{Luminosity function of thick disk and halo M dwarfs}
\label{sec:m_dwarfs}

Over $70\%$ of the stars observed by DES will belong to the thick disk and halo of the Milky Way. 
Low-mass main sequence (MS) stars with $0.08 < M/\Msolar < 0.50$ will dominate the DES stellar sample. 
About $7.5 \times 10^7$ red dwarfs are expected to be imaged ~\citep{Rossetto2011}.
Even in the presence of significant sample incompleteness and galaxy contamination, the number of stars in the typical colour range of M dwarfs will constrain the slope of the stellar IMF, $d\log N(m) / d\log (m)$, for these Galactic components to a precision of better than 0.1 dex.

The relatively straightforward colour selection for low-mass stars coupled with the large survey footprint of DES will allow us to study the structure of the Galactic disk with unprecedented detail. 
In particular, we will be able to detect new stellar substructures close to the Galactic plane and to search for asymmetries with respect to the northern hemisphere (comparing to similar star counts from SDSS, for instance)~\citep{Juric2008}.

\subsection{The luminosity function of thick disk and halo white dwarfs}
\label{sec:thick_disk}

White dwarfs (WDs) are the end point of stellar evolution for the majority of stars. The faint end of the WD luminosity function (WDLF) is occupied by the oldest members of a given stellar population, those that are the furthest along the WD cooling sequence. Therefore, the WDLF of the disk and the halo provides a method of dating the early star formation in each of these Galactic components \citep{Althaus2010}. Additionally, modelling of the WDLF can be used to constrain the stellar IMF and SFH \citep{Oirschot2014}.
WD are also an important class of objects for photometric calibration.
Both DES and other future large-area surveys will benefit from the discovery and characterization of WDs in the DES footprint.

\begin{figure*}
\begin{center}
\includegraphics[width=\textwidth]{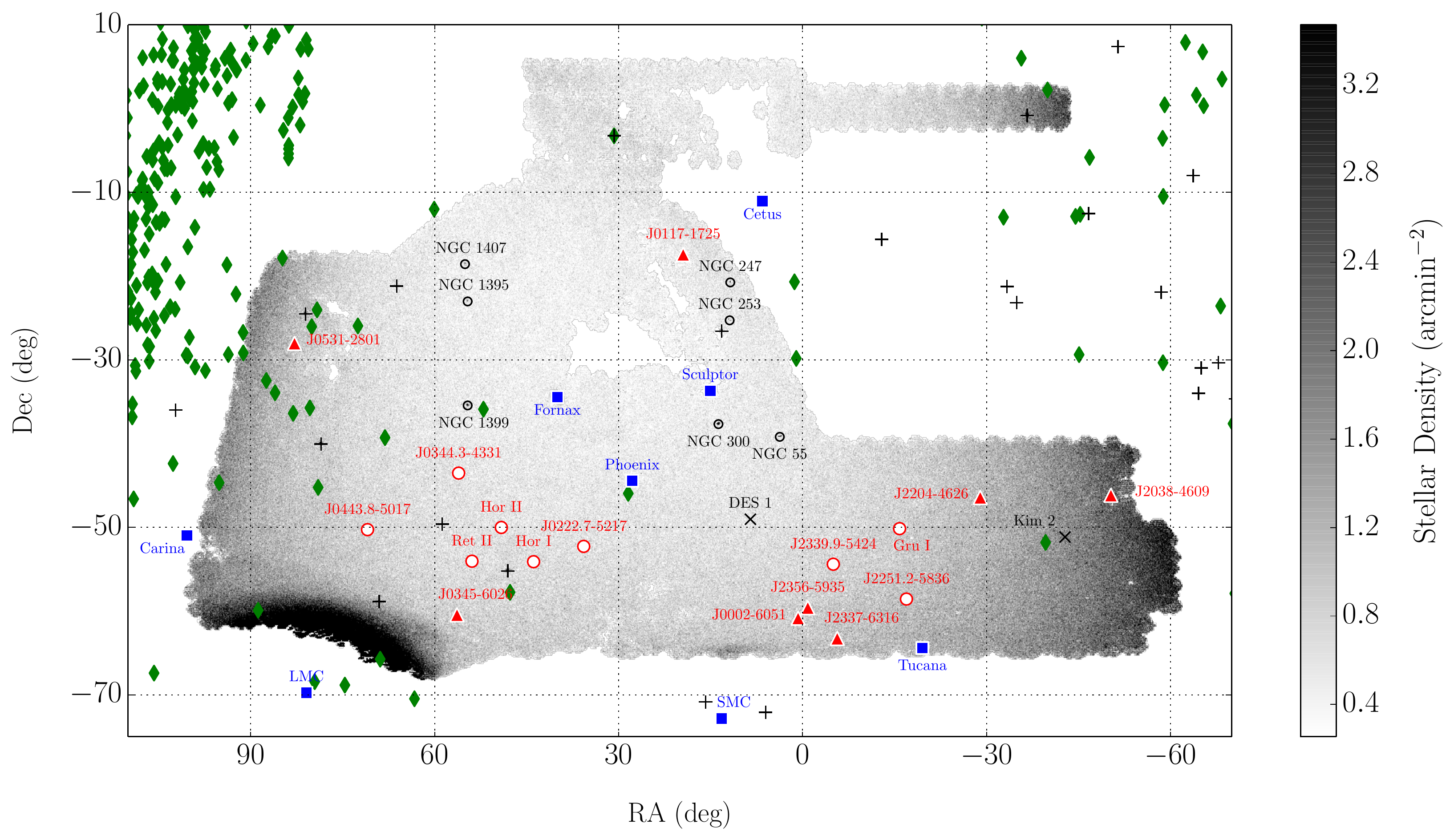}
\caption{
Cartesian projection of the density of stars observed in both $g$- and
$r$-bands with $g < 23$ and $g - r < 1$ over the DES Y2Q1 footprint
($\roughly 5000\deg^2$).
Globular clusters are marked with ``$+$'' symbols
\citet[][2010 edition]{Harris1996},
 two faint outer halo clusters are marked with
``$\times$'' symbols
\citep{Luque2015},
Local
Group galaxies known prior to DES are marked with blue squares
\citep{McConnachie2012},
dwarf galaxy candidates discovered in Y1
DES data are marked with red outlined circles, while the Y2 stellar
systems are marked with red triangles.
Open clusters from the WEBDA database are shown as green diamonds.
The periphery of the LMC can be seen in the southeast corner of the
footprint, while the Galactic stellar disk can be seen on the eastern
and western edges.
This figure was adapted from 
\citet{des_adw_2015}.
}
\label{fig6}
\end{center}
\end{figure*}

The search for nearby WDs will benefit greatly from the availability of proper motions. 
Since DES will not measure \band{\emph{u}}-band magnitudes, WD samples selected by \colourp{\emph{g}}{\emph{r}} colour will be contaminated by A and F stars from the Galactic disk and halo main sequence turn off (MSTO). 
The DES \band{\emph{Y}}-band images may be useful due to the signature of strong Paschen absorption lines in the spectra of DA WDs.
An excess of objects relative to model predictions for MS stars of similar colours may also be used to constrain statistically the WDLF. 

Unfortunately, MS contamination is aggravated for WDs close to the end of the cooling sequence, which are the most useful to determine the down-turn of the WDLF and estimate the age of each Galactic component. 
For this population, proper motion measurements will be crucial to separate kinematically a local WD population from more distant MS stars of similar colours.

\subsection{Membership and luminosity functions of Galactic open clusters}
\label{sec:clusters}

Galactic open clusters are unique astrophysical laboratories for understanding stellar evolution. 
Their ages can be fairly accurately determined by isochrone fitting of cluster colour-magnitude diagrams \citep[\eg,][]{Cargile2010a}, by determining the lithium burning-limit age \citep[\eg][]{Basri1996,Cargile2010b}, or by populating rotation-period \vs photometric colour space (gyrochronology) \citep{Barnes2003,Barnes2007,James2010}. 
In Figure~\ref{fig6},
 we plot the positions of open clusters from the extensive, although surely incomplete, WEBDA open cluster database (circa February 2015).\footnote{\url{http://www.univie.ac.at/webda/navigation.html}} We also plot the positions of globular clusters and Local Group dwarf galaxies, which will be discussed in \S \ref{sec:globulars} and \S \ref{sec:satellites}, respectively.
Seventeen of these open clusters are located in the DES footprint.
Only three of these seventeen clusters have well-determined physical properties, tending to be relatively nearby ($\ll 1 \kpc$), young ($\ll 4 \Gyr$), and either thin- or thick-disk ($z < 500\pc$) systems.
The well-studied Blanco 1 open cluster lies on the boundary of the footprint, and  was surveyed in $\band{\emph{griz}}$ during SV (to a $5\sigma$ limiting magnitude of $\band{\emph{g}} \simeq 22$).

It is relatively straightforward to establish membership using colour-magnitude and colour-colour diagrams for an exemplar open cluster, such as the rich, nearby ($120 \pc$), and young ($125 \Myr$) Pleiades \citep{Robichon1999, Stauffer1998}.
However, based on SV data, all higher-mass stars (earlier than G0V) are saturated in low-extinction DES fields out to $3 \kpc$ (which if perpendicular to the Galactic plane, is about three times the thick-disk scale height). 
At that distance (neglecting interstellar extinction), all cluster G, K, and M-dwarfs are detected ($S/N > 10$), whereas all sub-stellar objects (brown dwarfs) and degenerate compact objects (white dwarfs) are far below the detection threshold (see also Table 2, \citealt{Rossetto2011}).

For a closer cluster located at $\roughly 1 \kpc$ (about three times the thin-disk, and one thick-disk, scale heights), only K- and M-dwarf stellar members are non-saturated, although the entire brown dwarf population will still fall below the survey's magnitude limit. 
Of the seventeen open clusters in the DES footprint, three have well-determined distances and are closer than $300\pc$. 
For these systems, DES survey detections will only be possible by using their early-mid L-dwarfs and some post-Luminosity Function (LF) peak white dwarfs (depending on cooling age). 

In all likelihood, the primary utility of multi-filter photometry of open clusters in the DES footprint will centre on defining the number, space density, and photometric properties of very red low-mass stars and high-mass brown dwarfs. 
By constructing cluster mass-functions at the stellar/sub-stellar boundary, we will directly complement Galactic field IMF studies for late-M, L- and T-dwarfs. 
A side-by-side comparison of Galactic field and open cluster IMFs provides an empirical test for the magnitude and timescale of mass segregation and cluster disruption/evaporation \citep[\eg,][]{Moraux2004}. 
Further science goals arising from DES open cluster observations could include (a) detection and characterization of the WD sequence for the closest clusters, allowing an age derivation from theoretical cooling sequences to be calculated and compared to traditional age determination methods \citep[\eg,][]{Kalirai2001} and (b) sensitivity to magnetic activity-induced stellar variability for the most distant clusters. 
While the wide-field DES survey will possess too few epochs for precise determinations of rotational periods, cluster members will be able to be identified from their photometric variability \citep[\eg,][]{James2010}. 

\subsection{The shape of the Galactic stellar halo} 
\label{sec:halo}
There are discrepant estimates of the axis ratio (and its variation with Galactocentric distance) for oblate models of the Galactic stellar halo~\citep{Juric2008, Du2008, deJong2010}. 
Using the integrated numbers of stars in the typical colour range of old low-metallicity F and G stars, \citet{Rossetto2011} show that the axis ratio of the stellar halo can be constrained to within 1\% to 2\%, again assuming efficient star-galaxy separation.
More interesting is the prospect of exploring other halo geometries, including effects of triaxiality and asymmetries. 
The task of constraining these more complex models will greatly benefit from defining a joint DES/SDSS stellar sample.

\subsection{Detection and characterization of variable stars}
\label{sec:variables}

In addition to a large spatial footprint, DES will span a
five-year temporal period.  Each of the wide-field pointings will be
observed ten times in each band over the duration of the survey.  At
the same time, each of the ten DES supernova fields (covering a total
of $27$ sq deg) will be observed regularly with a seven-day
cadence.  This large multi-epoch data set presents an exciting
opportunity for discovering temporally variable stars.

Even the limited number of observations available in the wide-field
survey should be sufficient to discover new pulsating variable stars.
One straightforward method for identifying variable stars is to
compare the flux measured in each epoch to the median flux of the
object.  Preliminary searches for $\delta$-Scuti and RR Lyrae stars in
the SDSS Stripe-82 region show that this type of selection can yield a
variable star completeness of $\geq 70\%$ with a purity of $\geq
50\%$.  The completeness and purity of this technique is significantly
enhanced in the supernova fields where the number of observations is
much larger.  Additional calibration is possible using the known
populations of variable stars in the LMC and the Reticulum globular
cluster (Section \ref{sec:globulars}). 

\subsection{Detection and characterization of Galactic stellar streams}
\label{sec:streams}

Stellar streams often span tens of degrees on the sky; they originate from tidal stripping of stars from globular clusters and dwarf galaxies. Models of cold stellar streams yield constraints on the total mass distribution and on the number of dark matter subhaloes orbiting the Galaxy \citep{Yoon2011}, and also allow the inference of Galactic parameters and of the properties of the progenitor system \citep{Koposov2010,Kupper2015}.

Several known tidal streams, including one associated with the
Sagittarius dwarf galaxy, are expected to extend into the DES
footprint.  Often, these halo substructures are found by inspecting
the spatial distribution of halo tracers, such as horizontal branch,
red giant, or halo MSTO stars \citep[\eg,][]{Yanny2000,Belokurov2006}.
Indeed, one new stream has been discovered in the Y1 data
\citep{Balbinot2015b}. Another stellar stream associated with the LMC has been identified by \citet{Mackey2015}.

DES will also be sensitive to low-density extra-tidal features around
globular clusters.  A tool for identifying tidal features has been
developed within the DES collaboration and applied to NGC 2298, a
relatively low-latitude globular cluster~\citep{Balbinot2011}.  This
tool applies a matched filter algorithm to measure the density of
stars associated with a known cluster across the field. 

\subsection{Local Group globular clusters} 
\label{sec:globulars}

Local Group globular clusters provide a rich population of resolved
objects for study with DES.  These objects trace some of the oldest
and most nearby stellar populations.  They allow us to probe the
gravitational potential of the Local Group both through the bulk
orbital motion and through the tidal disruption of their member stars
(as discussed in \S \ref{sec:streams}).  Additionally, the dense
resolved stellar populations of globular clusters provide a testbed
for star-galaxy classification, determinations of proper motion, and
precision astrometric calibrations.  Globular clusters can also be
used to calibrate a photometric metallicity scale for DES filters in
the range of $-2.1 \leq {\rm [Fe/H]} \leq
-1.2$.\footnote{Complementing DES with \band{$u$}-band data will be
important for this task, since CN and NH features affect the
\band{$u$} filter \citep[\eg,][]{Ivezic2008, Marino2008,
Carretta2009}.}  Two globular clusters of particular interest to
the DES survey are \omegacen and Reticulum.

\omegacen is a dense Milky Way globular cluster located at $(\ra, \dec)=(13^{h}27^{m}, -47^{\circ} 29')$. 
The retrograde orbit of \omegacen suggests that this globular cluster may have originated as a dwarf galaxy and was tidally stripped by the Milky Way \citep{Carraro2000}.
Recent searches for tidal tails of Omega Centauri show no current tidal disruption \citep{DaCosta2008}; however, the wide coverage of DES and its depth at red wavelengths should provide unprecedented sensitivity to tidal features in the old stellar population of \omegacen.
In addition, the relative proximity and high stellar density of \omegacen makes it an excellent candidate for studying stellar proper motion.

The Reticulum globular cluster, located at $(\ra,\dec)=(04^h36^m,-58^{\circ}52')$, resides within the SV SPT-E field. 
Reticulum is at a distance comparable to that of the LMC and is likely gravitationally bound to that galaxy \citep{Walker1992,Dallora2004,Sollima2008}.
Its age and metallicity have been determined using HST-ACS data by \citet{Mackey2004} and are typical of the LMC old cluster population. 
The DES coverage around this cluster will allow the detection of extra-tidal features associated with this system, which in turn will help constrain the gravitational potential in the outer regions of the LMC.
The combination of spatial coverage and photometric depth will allow a more thorough investigation of the cluster's structure, with improved determinations of its current mass function, mass segregation, and binary fraction.
These characteristics will allow us to diagnose the role played by internal dynamical effects.

The Reticulum cluster serves as a distance reference, since it is known to be associated with the LMC and to contain many RR Lyrae stars \citep{Kuehn2013}. 
The wide-area coverage of the SPT-E data will help increase the census of these objects, specifically in the cluster outskirts.

Besides the two exemplars discussed above, \citet[][2010 edition]{Harris1996} lists eight additional Milky Way globular clusters that are covered by DES and several other globular clusters that lie close enough to the survey border to allow for searches for extra-tidal features (Figure~\ref{fig6}).

\begin{figure*}
\begin{center}
\begin{minipage}[b]{0.98\linewidth}
 \centering
  $\vcenter{\hbox{\includegraphics[scale=0.3]{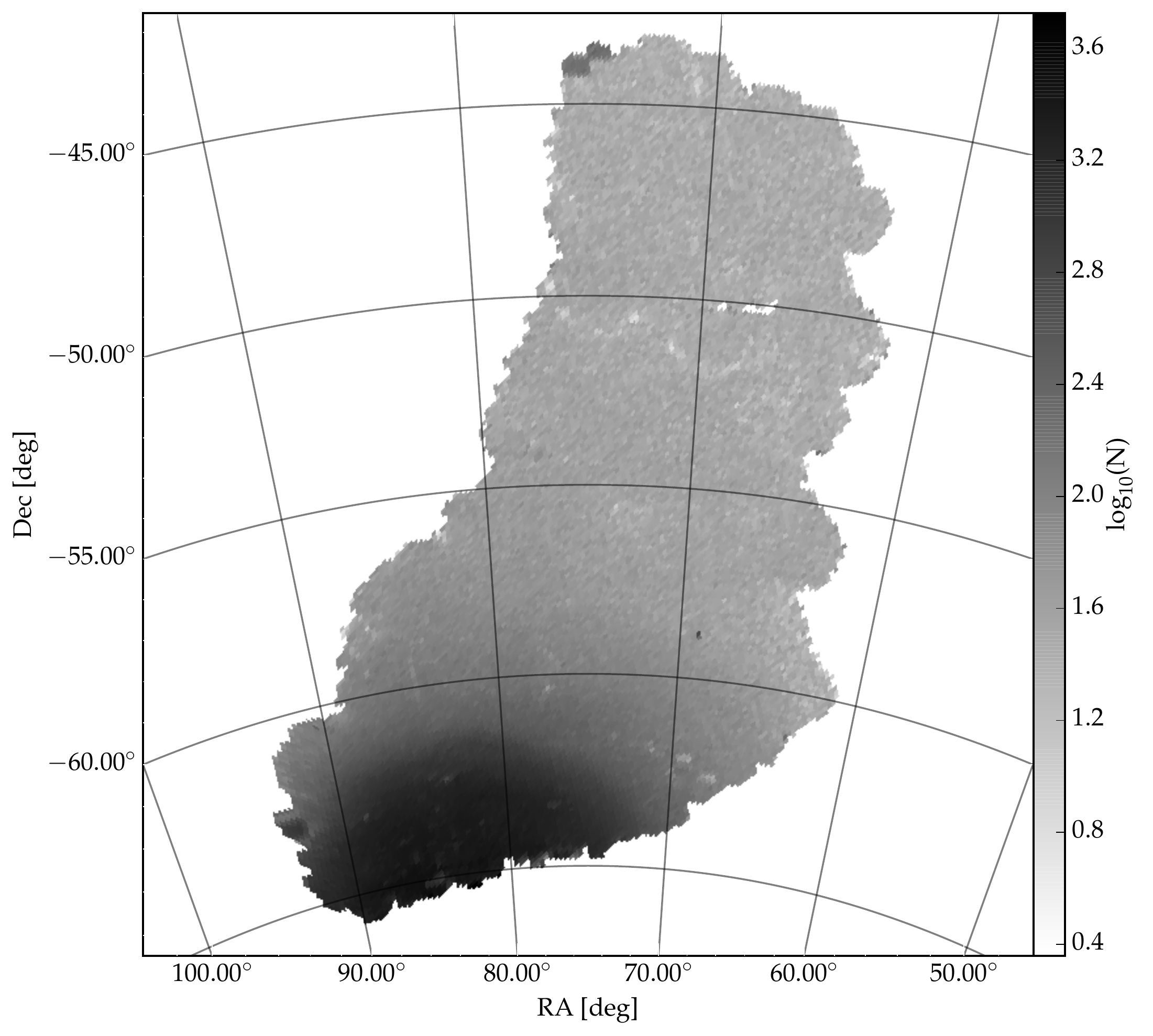}}}$
  \hspace*{.2in}
  $\vcenter{\hbox{\includegraphics[scale=0.3]{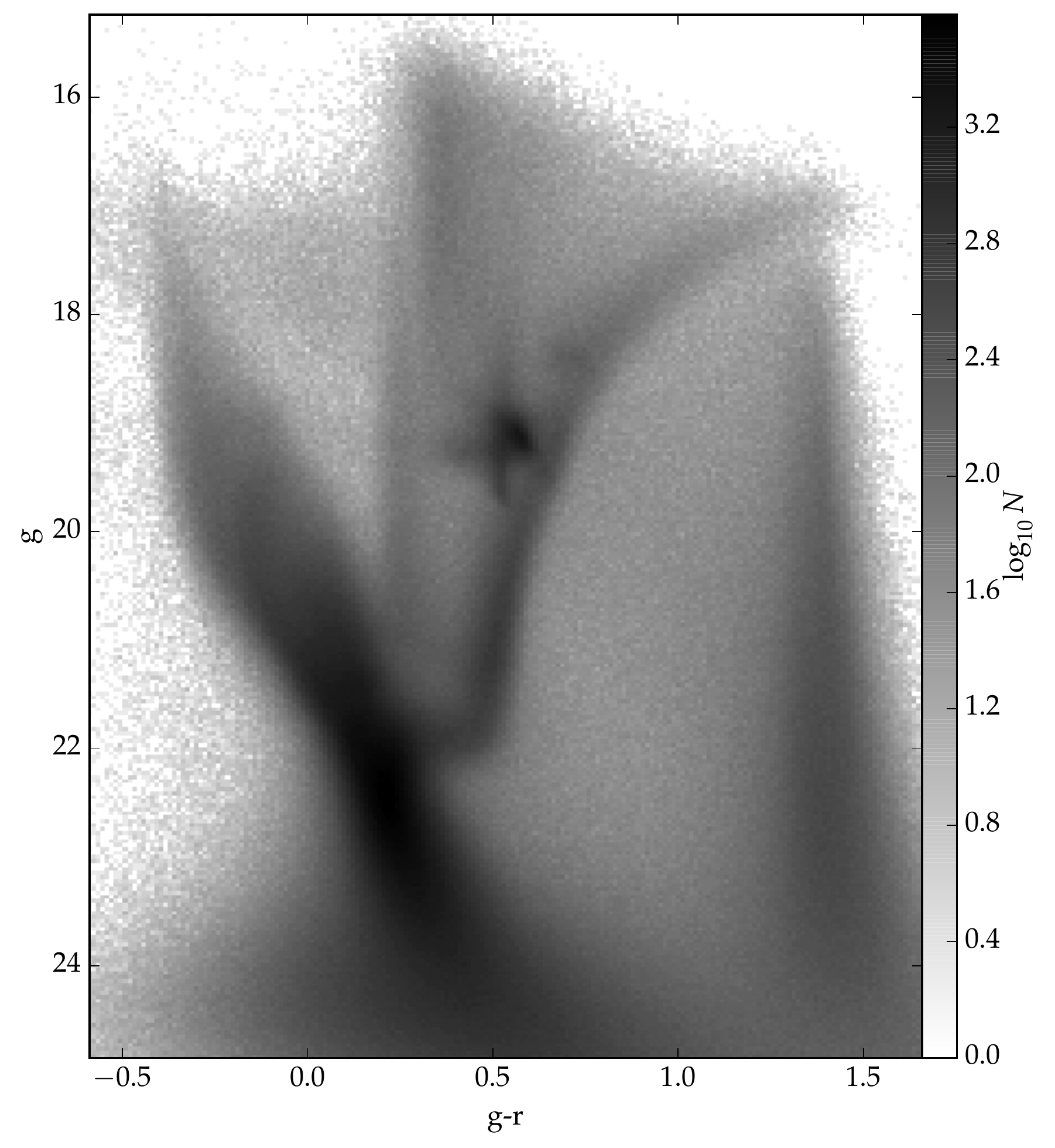}}}$
\caption{{\it Left:} Density map of sources classified as stars in the SPT-E 
region covered by the DES SV data; {\it Right:} $g~vs.~(g-r)$ Hess diagram of
all objects classified as stars and located  at $ DEC <  -55^o$. The
LMC main sequence stars (MS) are visible over almost the entire magnitude range. 
More than one main-sequence turn-off point (MSTO) is present, attesting
a large age range which includes an old population with MSTO at 
$(g,g-r) \simeq 22.5, 0.3$. Clear sub-giant and red giant branches (SGB and
RGB, respectively) associated with the older populations are also
seen redwards of $(g-r) \simeq 0.4$ for $g < 22.3$. The width both at the
main sequence and at the RGB attest a significant metallicity spread as well.
The red clump and horizontal branches are also clearly visible.
The AGB bump is also visible at $g=18.5$ and $g-r=0.7$.
}
\label{fig:lmccmd}
\end{minipage}
\end{center}
\end{figure*}

\subsection{Stellar structure at the outskirts of the Large Magellanic Cloud} 
\label{sec:lmc}

The largest contiguous area of the DES SV data is the SPT-E.  The
SPT-E field covers a region roughly $8^{\circ}$ on a side in the
declination range $-64^{\circ} \leq \dec \leq -46^{\circ}$.  The
southern border of the SPT-E region is located $\simeq 4^{\circ}$ from
the centre of the Large Magellanic Cloud (LMC).  Therefore, SPT-E
provides a uniform and deep sample of LMC stars ranging from $5 \kpc$
to $22 \kpc$ in projected distance from the LMC centre.
The left panel of \figref{lmccmd} shows the number density of stellar objects in the SPT-E field, where one can clearly see the steep rise in stellar density. The right panel shows a colour-magnitude diagram of a region $0.5$ sq deg located $\simeq 6^{\circ}$ from the LMC centre $(\ra,\dec)=(78^\circ06',-63^\circ54')$.
This region is clearly dominated by LMC stars spanning a wide range of ages and metallicities. 
The complex structure of He burning stars in the Red Clump and Horizontal Branch is clearly visible, including the secondary clump stretching vertically from the blue end of the main Red Clump \citep{Girardi1999}.

The contiguous coverage of the SPT-E field has already allowed us to constrain the structure and density distribution of the LMC disk out to $\simeq 20$ kpc. \citet{Balbinot2015} have fit the LMC star counts to a disk model and found it to have an inclination $i=38.7^{\circ}$ relative to the plane of the sky with a position angle of $\theta_0 = 129^{\circ}$ and an exponential scale radius $R_s = 1.11$ kpc. A significant variation in this latter parameter has been observed as a function of age: $\tau \leq 2\Gyr$ stars have  $R_s = 0.7$ kpc, whereas older stars are spread over twice this value along the LMC disk plane. The authors have also found this disk to be warped and likely flared in the outermost LMC regions probed. These results confirm, out to larger LMC radii, previously observed age and metallicity gradients with older and more metal poor stars located in the outer regions \citep{Carrera2008,Gallart2008}. 

The existence of a spherical stellar halo associated with the LMC remains an open question~\citep{Gallart2004,Carrera2011}. 
Determining the extent out to which LMC stars are to be found and whether this galaxy has its own system of satellite galaxies and other substructure also requires a large coverage in area and depth. Data from DES and from the Survey of the Magellanic Stellar History \citep[SMASH;][]{Martin2015}, when jointly analyzed, should provide a much more detailed picture of the stellar content of the LMC, the SMC, and the entire Magellanic system. These data will be sensitive to spatially varying SFH of LMC stars using the available CMDs at different directions across the field. Mapping the SFH of the LMC at large distances will allow a more detailed assessment of the radial extent of the observed age and abundance gradients, which suggest an outside-in galaxy formation history.

A more complete census of RR Lyrae stars belonging to the LMC and SMC is expected to result from these data, especially in its outermost regions. 
This will, in turn, provide an independent assessment of the structure and geometry of the external components of these galaxies \citep{Kinman1991}.
Finally, new star clusters and stellar substructure associated with LMC have been identified and characterized with SV data \citep{Pieres2015}. 

\begin{figure}
\begin{center}
\includegraphics[scale=0.45]{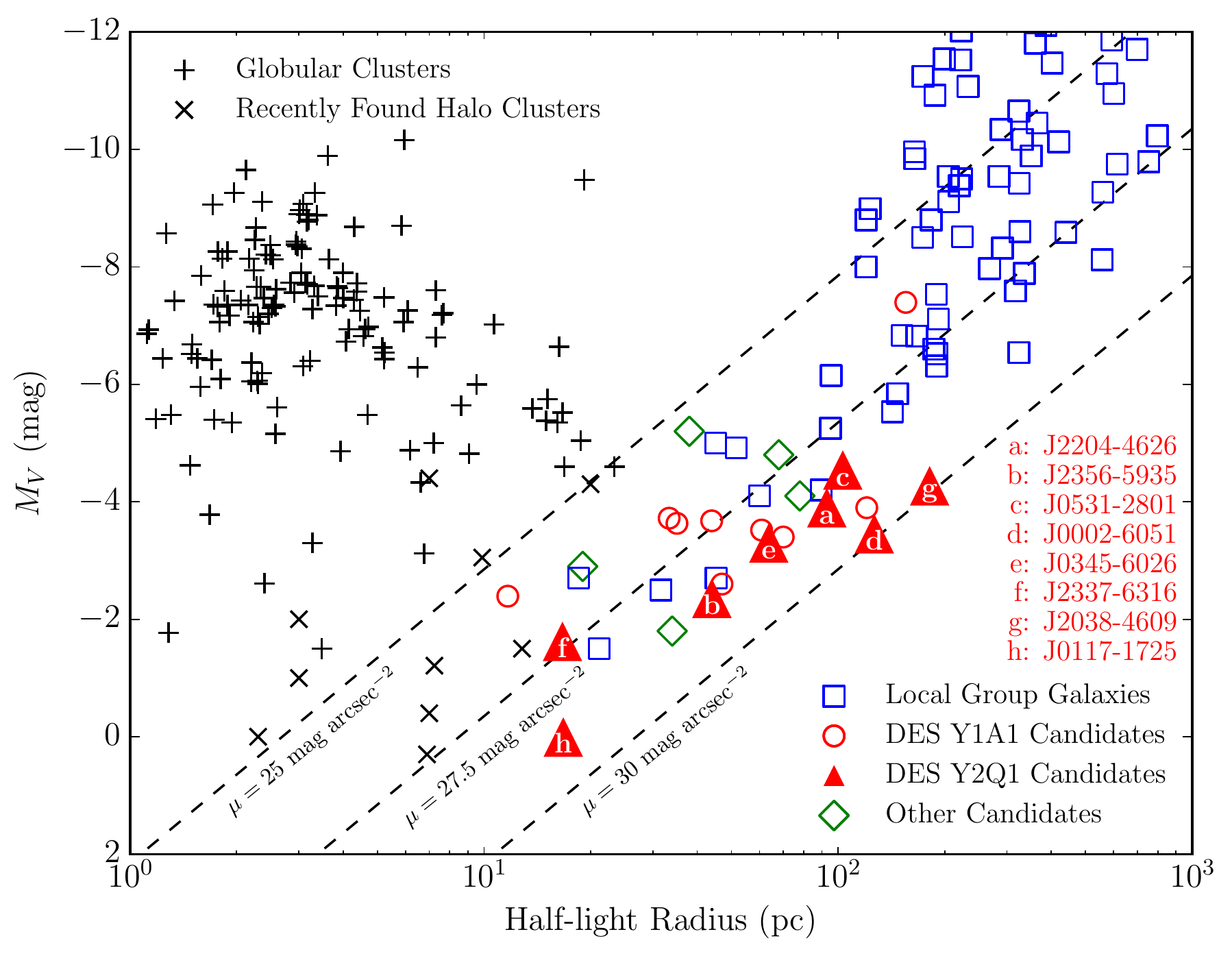}
\caption{Absolute \band{V}-band magnitudes and half-light radii of MW satellites and Local Group dwarf galaxies. MW globular clusters are marked with ``+'', fainter clusters and ambiguous objects are marked with ``x'', while spectroscopically confirmed dwarf galaxies are marked in blue. The red markers show the new dwarf galaxy candidates found in DES Y1+Y2, while green markers show candidates found in other surveys. Figure taken from \citet{Drlica-Wagner2015}.}
\label{fig:sizelum}
\end{center}
\end{figure}

\subsection{Detection and characterization of Milky Way satellite galaxies}
\label{sec:satellites}
\citet{Hargis2014} predict that three to thirteen new dwarf galaxies with $L > 10^3 L_\odot $ should be found by DES and a factor of a few as many below this luminosity. 
On the other hand, \citet{Tollerud2008} estimate that 19 to 37 new Milky
Way satellites could be discovered by DES, assuming that the
detectability of these satellites is similar to those
discovered in SDSS~\citep{Koposov2008, Walsh2009}, but extending to
the deeper limiting magnitude of DES.  However, \citet{Rossetto2011}
used an analytical model for the stellar IMF and the mass-luminosity
relation of MS stars to show that the number density contrast of
satellite stars relative to the background may increase by as much as
70\% in the DES sample when compared to SDSS.  Since DES will reach
the MS of old stellar populations out to $\simeq 120\kpc$, it will be sensitive to stellar systems at larger distances than SDSS. The DES Collaboration has developed a suite of tools, including matched-filter and maximum-likelihood algorithms, to search for Milky Way satellites in the DES footprint.

Analysis of Y1 DES data has already revealed nine new MW
satellite galaxy candidates that have sizes and luminosities
consistent with ultra faint dSph galaxies \citep{Bechtol2015,Koposov2015, Kim2015b}. Analysis of Y2 data added eight
more systems to this tally \citep{des_adw_2015}. 
In \figref{sizelum}, we compare the newly found
objects (red markers) to star clusters and other dwarf galaxies
from the Local Group. 

Since dwarf galaxies are very dark matter (DM) rich, they are
potential astrophysical targets for the indirect detection and
characterization of DM particles through annihilation into high energy
photons \citep{Bringmann2012}. Analyzing $\gamma$-ray data coincident
with the newly discovered dwarf galaxy candidates has been the subject
of much recent interest
\citep{Drlica-Wagner2015,GeringerSameth2015,Hooper2015}.  Meanwhile,
better estimates of the dark matter content of the new systems is
being actively pursued via spectroscopic observations
\citep{Simon2015,Walker2015,Koposov2015b}.  Gamma-ray observations of
new faint dwarf galaxies are expected to be sensitive to the generic
prediction for an annihilation signal from one of the prime DM
candidates, the weakly interacting massive particle (WIMP).

DES will also cover several classical Milky-Way satellites and Local
Group dwarf galaxies, including Fornax, Phoenix, Sculptor, and Tucana.
The deep, wide and homogeneous coverage
provided by DES will allow the spatial stellar distribution of these
systems to be probed beyond their tidal radii, therefore improving
their mass estimates and helping constrain the mass of the MW halo out
to their location.



 
\section{Search for optical counterparts of gravitational wave events}
The gravitational wave (GW) ground-based interferometric detectors, 
LIGO and Virgo, anticipate first detections within this decade 
\citep{2013arXiv1304.0670L}. 
Detectable GW events are expected from mergers of neutron star 
(or black hole) binaries out to about 200 Mpc. Such mergers are also expected to produce 
a fast, red, faint transient that optical telescopes might detect
\citep{2010CQGra..27q3001A,2014ApJS..211....7A}.  
This poses a unique opportunity to perform 
the first detection of gravitational and electromagnetic radiation from 
the same astrophysical sources and, therefore, to learn more
about the astrophysics
of neutron stars, the nucleosynthesis of heavy elements in the Universe, 
and its recent  expansion history. 
The DES Gravitational Waves (DESGW) project aims to
search for the optical counterparts of the first LIGO--Virgo detections
using DECam. The main challenge is to cover the large
regions where the source might be located: GW detectors have 
large uncertainties on the 
sky localization of a source (typically 300 sq deg  in the late 2015-early 2016 observing run, expected to improve by a factor of a few in the next run). 
Fortunately, DES is well-equipped to meet  
this challenge, with a 3 sq deg
field of view camera mounted on a 4 metre telescope.  

\begin{figure}
\includegraphics[width=\linewidth,trim={0 0 0 -50},clip]{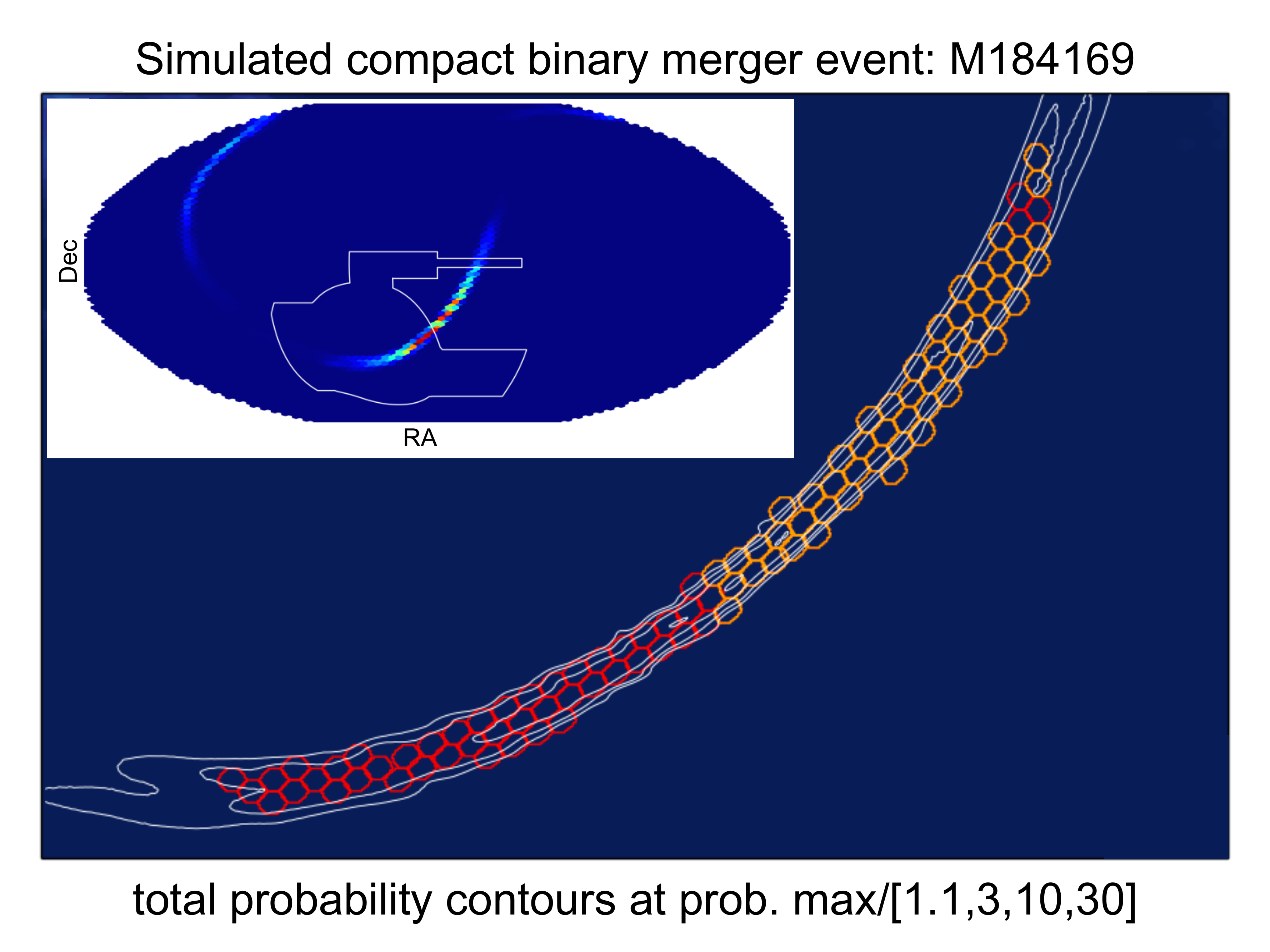}
\caption{Example of a simulated GW event trigger provided by the 
Ligo--Virgo Collaboration (LVC). Such simulations are used to optimize observing strategies and estimate detection probabilities.
The event M184169 shown in this example is at 119 Mpc distance and the LVC 
probability map shows the probability of the event falling in a given sky
pixel.  The inset shows the full-sky probability distribution derived
from the GW data, with the larger figure showing the area that would
be observed by DECam on the first night after the trigger
The red hexes are inside the DES footprint, the orange ones are outside.
The white contours represent the total LVC probability density
times detection probability, assuming a source of absolute magnitude
$M=-11$ at the trigger distance.  
}
\label{gwfig}
\end{figure}

The DESGW programme has been granted 3 additional nights of DECam time in 2015B semester
to perform the first search campaign and we expect to expand this programme 
in future years. The 3 nights of the DESGW programme are merged with the regular 
DES time allocation. When an event followup is to occur, we modify the 
observing plan for that night to perform observations covering the event's
region of interest and subtract the time used from our 3-night allocation. 

We have developed a system that watches for alerts from the GW detection consortium. Once an event trigger happens, the system processes the 
GW probability maps to 
determine the optimal observing plan for that particular event and estimates 
the probability of detection of the hypothesized optical counterpart. 
Figure \ref{gwfig} shows an example of a simulated GW event at 
a distance of about 119 Mpc. For this particular example, we assume 
that the absolute magnitude of the optical counterpart is $M=-11$ in 
$i$-band and that the light curve decays in about 10 days. 
These parameters are chosen to match the sensitivity range of the LIGO 
detectors in the late 2015-early 2016 run 
and the baseline 
emission model (merger of a pair of neutron stars with 1.4 solar masses each). 

For each event, we produce a plan for observations to be made 
within 24 hours from the trigger time.
The probability of detection takes into account the source model, the
distance, the depth achievable for that night (accounting for sky brightness, Galactic extinction and exposure times). 
We plan observations in $i$ and $z$ bands to make use of colour information. 
For the particular event shown in Figure~\ref{gwfig}, we expect to 
reach 49\% detection probability in 44  
hexagons
(each hex corresponds to 
the area of 1 camera field, which covers 3 square degrees).
This is a typical event. 
About 25\% of the simulations have total detection probability 
at this level or better. 
To perform these searches, we need to do difference imaging. DES does 
difference imaging routinely in 10 dedicated fields, the supernova fields.
We generalized the SN difference imaging pipeline to work on any area of the sky using any
available DECam image. This work will benefit other transient searches in DES
(e.g. solar system, near earth objects, nearby SNe, and quasars).

The start of operations of the LIGO--Virgo network of advanced gravitational
wave detectors combined with observations of their optical counterpart 
will potentially enable groundbreaking results, including:
the first detection of mergers of black holes and neutron stars, 
study of the evolution of binary systems, 
measurements of the equation of state of neutron stars,
determination of the origin of r-process elements in the Universe,
and a novel cosmological probe. 
In the long run, we expect that GW events will be used 
to measure the distance--redshift relation with great precision
\citep{1986Natur.323..310S,2005ApJ...629...15H}. A sample of 20-30 
well measured events can result in measurement of the Hubble parameter 
with uncertainty of 3\%. Future experiments may provide tens of thousands
of such events. 
The DESGW programme will capitalise on the 
groundbreaking potential of combined gravitational waves and optical
 observations. 
Simultaneous detection of GW and photons can also provide measurements of the speed of gravity and test modified theories of 
gravity \citep{Desai2008,Nishizawa2014}. 

 In February 2016 the LIGO collaboration reported the first detection of
gravitational waves, GW150914, resulting from the merger of two black holes
\citep{GWaves2016}.
DES provided optical follow up of this event, as decribed in 
\citet{Abbott:2016aa}, \citet{Annis:2016aa} and \citet{Soares-Santos:2016aa}.
No optical detections were detected, which is not surprising, as in the conventional model a binary black hole
merger is not expected to have any optical counterparts, and the DES
observations covered only part of the sky where the event was likely to
happen. However, DECam will be vital for future LIGO follow ups.


\section{Galaxy Evolution}

The DES data set, though designed for dark energy studies, presents a
goldmine for galaxy evolution science. The science covered is very
broad and includes studies of galaxies and galaxy properties across
redshift space from the nearby universe to epochs close to
reionisation around $z\sim 6$. This will be possible thanks to the
unprecedented combination of wavelength coverage, width, and depth of
DES+VHS photometry. Here we describe some of the broad science themes
enabled by the DES data: (1) the evolution of the multivariate and stellar
population properties of galaxies and their evolution with redshift,
(2) galaxy environment, (3) galaxy luminosity and mass functions and their
evolution with redshift, (4) massive galaxies at high redshift, (5) connecting
galaxies to their dark matter context, (6) galaxy properties in galaxy
clusters, and (7) intracluster light. A major shortcoming in traditional
galaxy surveys designed for galaxy evolution studies 
is the lack of sufficient volume required to overcome
limitations caused by both cosmic variance and the low number density
of massive galaxies. Thanks to the large survey area of 5,000 sq deg,
the volume sampled in DES is orders of magnitude larger than existing
galaxy evolution surveys, at the price of being shallower by several
magnitudes. With this unique combination of intermediate depth and
large volume, the DES data set will allow a major step
forward in the field. In the following we briefly discuss the key topics that we envisage being addressed by DES.

\subsection{Stellar population properties}
The relatively wide wavelength range covered by the DES photometry
allows far more than just the derivation of photometric redshifts, in
particular when the VHS near-IR data are included. The moderate depth even
in the reddest bands with 
$z\sim 23$ 
mag (AB) and $Y\sim 22$ mag will be
used for the analysis of the galaxy population for a relatively large
luminosity range up to redshifts $z\sim 1$. This particular epoch, at
which the Universe had about half of its current age, is critical for
understanding galaxy formation and evolution. It represents the
time of transition in the evolution of galaxies, where, viewed as a
function of increasing redshift, a largely passively evolving
population is believed to have broken up into its hierarchical pieces \citep[e.g.][]{Delucia2007}. The photometric bands $g$ to $K$ cover the
rest-frame wavelength range $2,200 < \lambda < 10,000 \AA $ around 
those redshifts, hence providing a good sampling of the spectral energy
distribution (SED). Stellar population model fits will provide key
galaxy parameters such as stellar population age, star formation
history, star formation rate, metallicity, dust attenuation, slope
of the initial mass function, absolute luminosity, and stellar mass
\citep[e.g.][]{Bolzonella2010,Maraston2010}. Redshift
evolution will help to break well-known degeneracies between some of
these parameters. The power of the DES data to study galaxy evolution
will largely lie in its ability to describe the joint distribution of a
wide range of galaxy properties, thereby determining correlations between galaxy properties and the environment, and how these correlations evolve
with redshift.  The joint distribution of several of these properties
is now well studied in the local universe from the SDSS \citep[e.g. ][]{BlantonMoustakas2009,Thomas2010}. DES will
allow extensions of these studies to $z\sim 1$ with a fine enough binning to
separate the various correlations to a degree that is not possible
with smaller surveys.

\subsection{Galaxy environment}
Thanks to the large volume sampled by DES, the data set will allow the investigation of galaxy environment beyond the local
universe up to $z\sim 1$. This large sample comes at the cost of
redshift precision, and this affects environmental measurements. Spectroscopic redshifts have been key in studies of
environmental effects in the local universe \citep{Thomas2010,Peng2010,Brough2013} and at higher redshifts (DEEP2, COSMOS); by contrast only photometric redshifts are available in DES.

In preparation for the DES, \citet{EtheringtonThomas2015} investigated
the impact of the photometric redshift precision on measures
of galaxy environment. They studied this by measuring environments using
SDSS spectroscopic and photometric
redshifts; they also simulated photometric redshifts with a range of
uncertainties. 
Photometric measurements  have a smaller dynamic range
than spectroscopic measurements because uncertain redshifts scatter
galaxies from dense environments into less dense environments. 
At the expected
redshift uncertainty of DES of $\sigma_z = 0.1$, they found a Spearman Rank
Correlation Coefficient of 0.4 between the photometric and
spectroscopic measurements.

\begin{figure*}
\begin{center}
\includegraphics[width=0.9\textwidth]{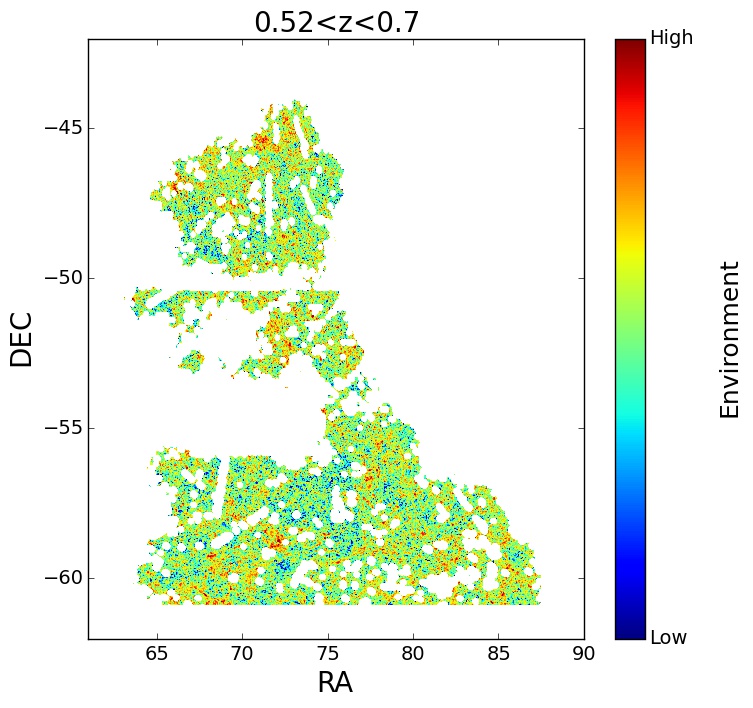}
\caption{Galaxy environment maps derived from photometric redshifts in
  a slice at redshift $0.52<z<0.70$ 
for a $\approx 100$ sq deg footprint overlapping with the SPT-E field
derived from the DES SV data. 
Dense environments are shown in red and sparse environments in blue.}
\label{fig:environment}
\end{center}
\end{figure*}

They further showed that the bivariate dependence of the
galaxy red fraction on stellar mass and environment \citep[e.g.][]{Peng2010}
 can be reproduced using photometric
redshifts. They concluded that photometric
samples with a redshift uncertainty of 0.1 must be approximately 6-16
times larger than spectroscopic samples to detect environment
correlations with equivalent fractional errors. This requirement is
easily met by DES.
Figure~\ref{fig:environment} maps environmental densities calculated 
from a subset of the 
SPT-E field in the DES SV data
using the full probability distributions of photometric
redshifts provided by the TPZ code \citep{Carrasco2013,Carrasco2014}.

\subsection{Luminosity and stellar mass functions of galaxies}

How the massive ($\gtrsim 10^{11}$ $M_{\odot}$) end of the stellar
mass function builds up over cosmic time is still a matter of debate.
 If constructed using hierarchical structure build-up, then this part of the mass function evolved significantly in the last $\sim 10\;$Gyr, during which time
galaxies accreted between 20 and 50 per cent of their final
mass. Observations, however, hint at a mass-dependent galaxy
number-density evolution with redshift, i.e. the larger the mass the
milder the number-density evolution \citep{Cimatti06,Marchesini09,Pozzetti10}. These discrepancies between observations and theoretical
models lead to uncertainties regarding galaxy formation. These uncertainties could be reduced
by having a survey with a volume large enough to detect a
representative sample of the most massive galaxies in the Universe
(the rarest part of the galaxy population), and deep enough to
identify a galaxy sample complete both out to a sufficiently low mass and out to $z\sim1$.

More recently, the Baryonic Oscillations Spectroscopic Survey \citep{Dawson12} provided a data set that allowed the study of the number
density evolution of massive galaxies up to $z\sim 0.7$. Based on this
data, \citet{Maraston2013} showed that the massive end of the galaxy
mass function has not been evolving since those redshifts \citep[see also][]{Bernardi2015}. DES will allow us to extend this study both to higher redshifts and to lower-mass galaxies covering key parameter space
of the galaxy mass function. Thanks to the large data set further key
parameters such as environmental density, galaxy type, colour etc. can
be folded in the analysis at high statistical significance.

\subsection{Galaxies at high redshifts}
The deep photometry and large volume probed by DES will further enable
statistical studies of galaxies at $z\sim 3-6$, and especially of the
massive end of the galaxy population where other surveys are limited
by small sample size. The overlapping VISTA data will be very useful for
these studies. The analysis of extended and statistically significant
samples of red galaxies will be pushed up to $z\sim 6$, close to the
epoch of reionisation.

Galaxies at $z\sim 3$ are usually singled out as `dropouts'
(a.k.a. Lyman-break galaxies; LBG) through successive passbands,
this makes use of the strong Lyman-break signature in their SEDs
\citep[e.g.][]{Steidel95}. The multi-band data from DES will be sufficiently
deep that they can be used to select galaxies from $z\sim 4$ to 6 in
three redshift bins as $g,r,i$-dropouts. If young, massive galaxies
have already assembled at high redshift, then DES will be able to
detect them. This observation is exciting as the presence of such
objects is expected from the fossil record of local massive galaxies
\citep{Thomas2010}, but could not be easily accommodated in theories
of hierarchical galaxy formation based on cold dark matter cosmogonies
\citep[e.g.][]{Baugh2005}. \citet{Davis2013} investigate the potential
for DES to detect such objects in a study based on stellar population
modelling and the DES simulated data. They show that very young massive
galaxies would be found, if they exist and are not heavily dust
obscured. The combination of sky coverage and photometric depth of DES
is the key. Narrow, deep surveys, usually designed for the detection
of high-$z$ galaxies \citep{Stanway2003, Bouwens2010}, do not cover
enough volume to find rare massive galaxies. DES will play a
unique role in this field as its 5000 sq deg survey area will enable us to set a stringent constraint on the brightest
end of the galaxy luminosity function between $z\sim 4$ and
6. A precise knowledge of the bright end behaviour of the luminosity
function will have an enormous impact on a series of critical
cosmological questions, such as the star formation process at very
early epochs and the reionisation of the Universe.

In addition to the searches for massive high-redshift galaxies in the DES wide survey, the DES deep supernova fields will also be a goldmine for studying the high-redshift universe and the Epoch of Reionization.  One of the most fundamental questions in galaxy formation research today is understanding when massive galaxies were first formed and quenched. Thousands of  high-redshift galaxies at $z\gtrsim6$ are now being routinely discovered using \textit{HST} \citep[e.g.][]{McLure:11, Bouwens:15}
, but the bright-end of the galaxy luminosity function at these redshifts can only be constrained using ground-based surveys over much wider fields. Recently \citet{Bowler:14}
 have shown that the $z\sim7$ luminosity function follows the dark matter halo mass function, which could mean that quenching mechanisms such as AGN feedback are yet to come into effect at these early epochs. Therefore $z\sim5-6$ represents a very interesting redshift range where the effects of mass related quenching first begin to be seen in the galaxy population. 

\begin{figure*}
\begin{center}
\begin{tabular}{ccc}
\large{DES r-band} & \large{DES i-band} & \large{DES z-band} \\
\includegraphics[scale=0.4]{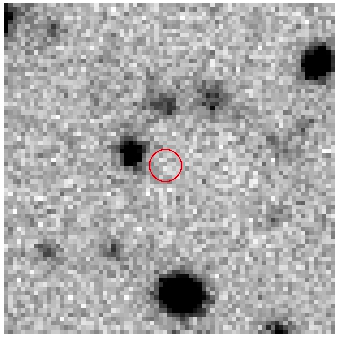} & \includegraphics[scale=0.4]{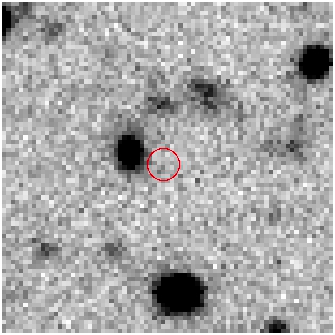} & \includegraphics[scale=0.4]{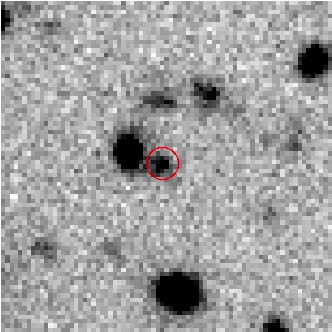} \\
\end{tabular}
\caption{DES $r$, $i$ and $z$-band images in the DES deep fields for a spectroscopically confirmed galaxy at $z=6.07$ galaxy from \citet{Willott:13}. The galaxy position is marked by the red circle in all three images. As can be seen, this high-redshift galaxy is undetected in the DES $r$ and $i$-bands but detected in the redder DES $z$-band. This characteristic ``drop-out" behaviour of high-redshift galaxies can be exploited to find similar $z>6$ galaxies in the DES deep fields.}
\label{fig:highz}
\end{center}
\end{figure*}

The DES deep fields will reach 10$\sigma$ co-add depths of $\sim$25-26 over $\sim$27  sq deg \citep{Bernstein2012} 
 enabling new  samples of luminous $z\sim5-6$ (LBGs) to be discovered. 
Figure~\ref{fig:highz}
shows a spectroscopically confirmed $z=6.07$ galaxy \citep{Willott:13}
 in the DES Year 1 data. As can be seen the galaxy is robustly detected in the DES $z$-band but undetected in the $r$ and $i$-bands. This highlights that the DES Year 1 data is already sensitive enough to detect $z\sim6$ galaxies and the characteristic ``drop-out" behaviour of high-redshift galaxies illustrated in Figure~\ref{fig:highz}
will be exploited to identify new high redshift galaxies in the survey data. The DES deep fields also overlap with deep infrared surveys using VISTA \citep{Jarvis:13}
 and \textit{Spitzer} \citep{Mauduit:12}.
Combining the optical DES data with these infrared datasets will also enable constraints on the stellar masses and spectral energy distributions of $z\sim5-6$ galaxies.

\subsection{Galaxy Structure and Morphology}

One of the primary ways in which galaxies are studied is through the analysis of their structures and morphologies.   The gross overall morphology of a galaxy (spiral, elliptical, peculiar) gives an indication of its formation state. Furthermore quantitative measures of structure allow us to determine the distribution of light in galaxies, including how concentrated its light is, and whether or not a recent dynamical or star formation event has occurred.   This is an area where DES will advance our knowledge considerably, given the higher resolution and depth of the data in comparison to previous surveys such as SDSS.  

The type of measurements that DES will carry out include non-parametric and parametric fitting of galaxy structure. This includes measuring the Concentration / Asymmetry / clumpinesS (CAS) parameters for the brighter sample of galaxies in DES \citep[e.g.][]{Conselice2003},
as well as using non-parametric fitting of surface brightness profiles to obtain effective radii, as well as Sersic indices.  These can be used together to trace the evolution in structure as a function of redshift, find galaxy mergers, and quantify the structures of galaxies as a function of environment.   A big mystery is why galaxies at a given mass become less concentrated and smaller at higher redshift \citep[e.g.][]{Buitrago2013},
 something that DES can address by examining how these features vary as a function of environment, star formation history and halo mass.   The overall morphologies can also be measured as part of Galaxy Zoo citizen science, whereby large members of the public can contribute to this aspect of DES.

The wide-field DES survey will image a total of nearly 2 million galaxies with stellar masses 
M$_{*} > 10^{10}$ M$_{\sun}$ at $z < 0.3$.  This is significantly more than any other survey to date, and will allow us to determine how the morphologies and structures of massive galaxies vary as a function of other parameters such as stellar mass, environment, and halo mass.  We will also have some limited ability to trace the time evolution of galaxy structure.  At increasing redshifts, a largely passively evolving population is believed to have broken up into its hierarchical pieces \citep[e.g.][]{Delucia2007},
 and the familiar Hubble-type morphologies give way to increasingly irregular type galaxies \citep{Mortlock2013}.
   DES will be able to trace the final stages of this process.

Within these measurements statistical studies of the environmental dependence of galaxy properties, such as the so-called morphology-density relation, can be carried out as a function of redshift, and on a larger scale than has been possible before.  There is a well known correlation between galaxy structure and environment, such that early type galaxies are more likely to be found in dense environments \citep[e.g.][]{Dressler1980}.
  With DES data we can go beyond simple morphologies and examine how the sizes and light concentrations of galaxies relate to environment and thereby tease out the effects of local and global densities on the structures of galaxies.

Furthermore, the DES data will allow us to study the faint galaxy population in the nearby universe for the first time, with several million galaxies fainter than M$_{\rm B} = -17$ being resolved.  This will give us the morphological properties of satellite galaxies, and allow us to determine the properties of dwarf systems which are found in void regions.  This relates to recently discovered galactic conformity, whereby the central galaxies and their satellites have similar colours and star formation histories \citep[e.g.][]{Hartley2015}.
   DES will for the first time allow this conformity to be studied as a function of morphology and structure as well as the colour.

\subsection{Evolution of galaxy properties in clusters}
There has been much progress on characterising the galaxy population
in clusters at lower redshift ($z < 0.3$) with SDSS
\citep{Hansen2009}, where excellent statistics have allowed selecting
samples that share many common properties (e.g. by colour, position,
and whether central or satellite) to isolate different contributions
to galaxy evolution from typical physical processes in clusters such
as ram pressure stripping, galaxy harassment, and cluster tidal field
interactions. Work with smaller volumes at higher redshift has
demonstrated that some of the observed trends seem to be in place at
$z\sim 1$ \citep{Gerke2007}, but the direct comparison of low and high
redshift samples is hampered by the very different selection functions
of the separate programmes. The rich and statistically significant
sample of galaxies in clusters in DES data will enable such studies to
push significantly further in redshift. Crucially, the DES data will
provide not only a large sample of cluster members, but also a large
sample of field galaxies with which to compare the cluster galaxies,
allowing for a self-consistent measurement of the effect of cluster
environment on galaxy evolution to $z\sim 1$. The DES data will enable
greatly improved constraints on the galaxy population as a function of
halo mass to high redshift, and, in comparison with large DES mock
catalogs from high-resolution simulations, will be a key to unravelling the physics of galaxy transformation in cluster environments \citep{Wetzel2012,Haines2015}.

The brightest cluster galaxies (BCGs) in galaxy clusters tend to obey
different relations between their mass, size and luminosity compared to
satellite cluster galaxies \citep{Bernardi2007, Vonderlinden2007}. The
properties of BCGs are closely coupled to host cluster mass
\citep[e.g.][]{Hansen2009}, as is expected from their privileged
position near the centre of the dark matter potential well of
galaxy clusters. Although the uniqueness of BCGs is often attributed
to their likely distinct merger histories compared to more average
galaxies \citep[e.g.][]{Delucia2007,Lidman2012}, observational
results remain inconclusive regarding which physical mechanisms are
actually at work \citep{Brough2008}. DES's statistically significant
sample of these rare objects will allow a systematic characterisation
of BCG properties as a function of host halo mass and redshift up to
at least $z\sim 1$ \citep{Zhang2015}.

\subsection{Measurement of the intracluster light}
Galaxy clusters are found to contain a population of stars that
are not bound to individual galaxies but which rather trace the cluster
potential. The resulting diffuse intracluster light (ICL), which may
contain 10-30\%\ of the total cluster light \citep{Zibetti2005,
  Gonzalez2005, Krick2006}, is a fossil remnant of galaxy evolution in
the cluster environment integrated over the course of a cluster's
accretion history. The change in ICL as a function of redshift would
provide valuable constraints on models of galaxy evolution in the
cluster environment \citep{Willman2004, Monaco2006, Conroy2007}. While
there has been detailed work to study a few tens of individual
clusters at low redshift \citep[e.g.][]{Gonzalez2005, Krick2006,
  Pierini2008}, constraints from these programmes remain weak owing to
the small samples available. ICL measurements on stacked samples of
clusters in the SDSS have been made \citep{Zibetti2005}, but uncertain
systematic effects of such a technique make the interpretation of
these results challenging. Based on current work using SDSS co-added
southern stripe data to measure the ICL of individual clusters, it is expected that DES data, with careful control of data reduction
tasks such as flat fielding and sky subtraction, will enable the study
of ICL in individual clusters to $z \sim 0.3$ over the full 5000
sq deg, an unprecedented volume for such work. Measurements from
stacked clusters may be possible to even higher redshifts.

\subsection {Galaxy biasing and halo modelling} 


The spatial distribution of galaxies does not necessarily follow the
distribution of mass fluctuations. This is commonly referred to as
`galaxy biasing'; it could be linear/non-linear, deterministic/stochastic, local/non-local,  and a function of scale and cosmic time. DES
has the power to constrain biasing both on its own and in combination with other data sets.  When deriving cosmological parameters from DES  biasing appears as a
set of ’nuisance’ parameters, but if the cosmology is known
(e.g. LCDM) or marginalised over, understanding biasing is important
in its own right  as a probe of galaxy formation and evolution \citep{Benson2000, GaztanagaScoccimarro2005, Zehavi2011}.

Galaxy biasing was recognised when it was noticed that different
populations of galaxies have different clustering strengths
 \citep{DavisGeller1976, Dressler1980}. A physical mechanism
for galaxy biasing, namely that galaxies would tend to form in peaks in the
matter density distribution thus being more clustered than the
underlying matter distribution, was suggested by \citet{Kaiser1984} and
developed by \citet{Bardeen1986}.  In this picture, more massive
(and thus rarer) tracers are naturally more highly biased.  Early
clustering measurements also indicated that galaxy bias cannot be
linear \citep{Gaztanaga1992, FryGaztanaga1993}. Indeed, a
linear bias relation could not be preserved through the non-linear
growth of structure, and analytic models \citep[e.g.][]{MoWhite1996} and
$N$-body simulations \citep[e.g.][]{GuoJing2009} provide a description of
this non-linearity.  Biasing is also likely to be stochastic \citep{Dekel1999} since it is not possible to specify the galaxy
distribution without also specifying other ‘hidden variables’ such as
luminosity, temperature, physical size etc., which would cause
some physical scatter in the relation between the galaxy and matter
density fields.  Additionally, stochasticity is introduced into the
measurement because of the discrete samples of the density field
selected, i.e. `shot noise'.  Galaxy biasing evolves with redshift
\citep{Nusser1994,Fry1996,Tegmark1998}, being
naturally larger at high redshift as the first galaxies to form would
have done so in the densest regions. It is also scale dependent at
small physical scales where the non-linear effects of galaxy formation
are important, although this is weak on large scales \citep{Mann1998}, 
being approximately scale invariant above 20-40 Mpc/$h$
 \citep{Crocce2013}.
 \citet{Clerkin2015} parameterised the redshift dependence of galaxy biasing.
Biasing can also be expressed as a more general expansion in functions
of the density field, and the related bias renormalization
 \citep{McDonald2009, Assassi2014}. Galaxy bias has also
been formulated in terms of the `effective field theory of LSS' \citep{Angulo2015}.


Another popular approach, motivated by our modern understanding of
galaxy formation in LCDM universes, is to parameterise the relationship
between galaxies and the dark matter distribution assuming that all galaxies live in dark matter haloes or
subhaloes, discussed e.g. in  \citet{Tinker2010} and earlier rerferences therein.
The connection between galaxies and dark matter haloes can be specified by the Halo Occupation Distribution (HOD), the probability distribution of the number of galaxies given the properties of the containing haloes, such as their masses, and given the relation between galaxy and dark matter spatial and velocity radial distributions. The HOD can also be studied for particular galaxy types, classified according to properties such as luminosity, stellar mass, colour, and star formation rate. At
low redshift this has been used with great success in SDSS, to the
extent of isolating which galaxy properties are correlated with each
other and with their host dark matter haloes, which in turn has
cleanly narrowed down the wide range of suggested causes of various
properties. At higher redshift, where the galaxies are less evolved and more gas rich, and where other energy inputs differ (e.g. there is more
quasar activity), the picture will likely be different, and sewing it
together with the low redshift universe will likely also lead to new
and interesting insight on galaxy properties and evolution. DES is
ideal to pursue such studies and to push previous results based on SDSS
to higher redshifts. Key constraints on the HOD of galaxies come from their clustering properties,
including the angular two and three point cross-correlation functions, as well as from
the galaxy-mass correlation function and the relationship of galaxies
to clusters and to underdense regions. The DES data will be
powerful in constraining halo model parameters such as the
occupation number and the density profile as well as determining
whether the halo occupation depends on halo properties other than the
halo mass.

In the context of DES there are several approaches to quantifying
biasing.  These include:
(i)  assuming an underlying model for the clustering of dark matter from an analytic model or simulations. The galaxy biasing can then be derived from DES clustering of galaxies (by type), e.g. from the respective two-point correlation functions: 
$ b^2 = \xi_{gg}/\xi_{mm}$. See application to SV data in 
 \citet{Crocce2015} 
(while biasing  was derived there for a magnitude limited galaxy sample, future work will also apply it to volume-limited samples). The measurement can be generalized for halo model parameters.
(ii) Using maps of dark matter itself.
The weak lensing from DES itself can be turned into
convergence (kappa) maps \citep{des_cc_2015, 2015arXiv150403002V},
which effectively is the integral of the dark matter fluctuations
along the line of sight (weighted by a lensing kernel).  
Biasing (for a magnitude limited sample) was derived recently by  \citep{Chang2016}.
Similarly,
mass maps were derived from the fluctuations of the Cosmic Microwave
Background by Planck  and SPT. Cross-correlations of the DES galaxies with such mass maps can
tell us the biasing \citep{Giannantonio2015}.  Alternatively it can
be derived from `counts in cells', so getting the biasing
$b = \delta_g/\delta_m$, `localised' over the redshift range probed by
the relevant lensing kernel.

DES will also have significant power to constrain the bias and the
relationship between galaxies and haloes through combinations of
various observables, including galaxy-galaxy lensing, the connection
between galaxies and clusters, and the connection between galaxies and
underdense regions.

\subsection {Galaxy Intrinsic Alignments}


Weak gravitational lensing is a key DES cosmology probe. It uses the
coherent distortion of galaxy shapes due to gravitational lensing by
large-scale structure, to measure both cosmic expansion history and
the growth of structure over time. Even without this lensing effect,
it is believed that intrinsic (i.e. unlensed) galaxy orientations and
shapes are aligned to some extent because neighbouring galaxies form
in the same large-scale gravitational potential. See
\citet{Troxel2014, Joachimi2015, Kiessling2015} and \citet{Kirk2015}
for recent reviews of the topic of galaxy alignments.

This \emph{intrinsic alignment} (IA) of galaxies acts as a contaminant
to weak lensing studies because the IA effect mimics the lensing
signal, meaning that the observed cosmic shear is the sum of both
contributions. Unless IAs are properly accounted for, any cosmological
constraints will suffer bias, often at a very significant level. In
the full DES survey, taken as a single dataset, the impact of IAs is
forecast to be important, contributing up to 10\% of total observed
cosmic shear at some scales. In the type of tomographic analysis that
produces useful cosmological constraints, IAs become critically
important; they are forecast to contribute over 50\% of the observed
cosmic shear signal in some tomographic bin pairs, at some scales
\citep{Joachimi2010}.

The intrinsic alignment of galaxy shapes is, however, much more than a
nuisance in the pursuit of the measurement of dark energy. Shape,
orientation or spin alignments can tell us a lot about the nature of
galaxy formation and evolution, as well as shedding light on the
relation between the baryonic component and its host dark matter haloes
and helping us to understand the morphology of the cosmic web. In
general, different physical alignment mechanisms are believed to be the
source of IAs in red and blue galaxies, while IAs may also vary as a
function of redshift, luminosity and other properties.

Most studies which set out to directly measure IAs
used a sample of galaxies with spectroscopic redshifts, of which at
least some subset had good shape measurements \citep{Hirata2007, Mandelbaum2013,Singh2015}. A common method is to measure
the two-point correlation of galaxy position with the component of
galaxy ellipticity along the line of sight connecting a galaxy pair,
$w_{\rm g+}$ \citep[e.g.][]{Singh2015}. The relative inaccuracy of the
photometric redshifts from DES make precise measurement of this
statistic challenging, though \citet{Joachimi2011} demonstrated that
it was possible to make measurements of large-scale IAs using data
with photometric-quality redshifts. They made a positive detection of
IAs in a sample of Luminous Red Galaxies (LRGs) with errors of $<10\%$ using the galaxy
position-ellipticity cross-correlation.
DES has derived very accurate photometric redshifts for red
galaxies using the redmapper \citet{Rykoff2016}  and redmagic samples, with scatter of $0.017(1+z)$ for
this sample; this will enable tighter constraints.  The photo-$z$
estimates are expected to be less accurate for other galaxy samples,
but even here DES offers the chance to push to redshifts deeper than
the $z < 0.7$ used in this previous work.
We could also employ tomographic galaxy-galaxy lensing \citep[e.g.][]{Blazek2012}
 to measure IA with the typical lensing sources.

\citet{Heymans2013} showed that a dedicated cosmic shear survey can
make informative indirect measurements of IAs. Here the authors
assumed a model for IAs, the non-linear alignment model of IAs
\citep{Bridle2007}, which has been shown to fit well at linear and
quasi-linear scales. Using two-point ellipticity correlation function
data from the CFHTLenS survey, \citet{Heymans2013} made a simultaneous
fit to cosmological parameters and their IA model, which they
parameterised by a single free amplitude parameter. Split by colour,
the analysis showed the IA of blue galaxies was consistent with zero,
while IAs in red galaxies were detected at close to $2\sigma$
confidence. Both results are consistent with theoretical
expectations. DES SV data are broadly comparable to CFHTLenS in terms
of sky coverage and depth. The full DES five year survey will increase the
available area by more than a factor of 30, making it realistic to
expect constraints on IA amplitude which are an order of magnitude
tighter than those from CFHTLenS.

This increased constraining power will also allow more ambitious
parameterisations of the IA model and the comparison of different
models. Of particular interest are measurements of the IA signal as a
function of luminosity, where a positive correlation has been
reported, and redshift, where no evidence for evolution has been found
\citep{Singh2015}. In particular, the deep redshift baseline of DES,
compared to the spectroscopic datasets mostly used in the past, is
likely to improve knowledge of IA evolution; this is of importance in
understanding the mechanisms which source galaxy shape alignments.

The DES dataset will be extremely useful for joint studies of IAs in
combination with data from other surveys. Overlapping spectroscopic
surveys (see Figure~\ref{fig:footprints}) 
combined with high-quality shape
measurements from DES will allow for high-precision IA measurements.
In the future, the Square Kilometere Array (SKA) large volume radio survey offers potentially
novel synergies, including the comparison of shape measurements made
in the optical and radio wavelengths \citep{Brown2014}, large numbers
of spectroscopic-quality redshifts from the HI survey
\citep{Abdalla2015}, and potentially the use of radio polarisation or
radial velocity measurements to directly access information about the
intrinsic galaxy ellipticity, allowing for the calibration of IAs in
optical surveys such as DES \citep{Whittaker2015}.
Cross-correlating lensing maps derived from the CMB (which are not affected by IA) with those from galaxy shapes measured in DES will allow a novel measurement of IAs \citep[e.g.][]{Kirk2015b}.

\section {Cluster properties}


In addition to their roles as cosmological probes, clusters of
galaxies offer important insights into a slew of astrophysical
processes including the formation and evolution of galaxies, the
effects of AGN feedback, tests of the nature of dark matter, the
acceleration of relativistic particles, and the amplification of
magnetic fields (among other astrophysical plasma processes). Their deep dark matter potential wells retain baryons, allowing multiple cold and hot phase baryonic components to be observed. They are
also powerful gravitational lenses offering probes of the mass and
structure of the clusters themselves as well as the magnification of
very high-redshift galaxies and quasars, allowing study of these very
distant objects.  DES will detect on the order of 100,000 clusters out
to $z\sim1$, comparable to the number of haloes with masses above $10^{14} \Msun$ expected in an LCDM universe over this sky area \citep{Allen2011}.

Overlap with the SPT provides around 1,000
clusters detected via the Sunyaev-Zel'dovich effect (SZ) between the
existing 2500 sq deg survey and a nearly complete SPTpol survey;
thousands more are expected from the upcoming SPT-3G survey.  Existing
XMM-Newton and Chandra data likewise give X-ray coverage of approximately
1000 clusters.  For these samples, DES provides both photometric
redshifts and cluster confirmation through the existence of galaxy
overdensities, and this overlap enables multiwavelength probes of
cluster and cluster galaxy evolution.

Clusters also act as signposts of the cosmic web, and both the
large-scale galaxy distribution and weak lensing shear signal from DES
data will allow the identification of supercluster and void regions
and filaments joining clusters \citep{2015arXiv150403002V}. This in turn will allow
tests of structure formation and galaxy evolution models 
as well as potentially identifying sightlines for searches for absorption
due to the warm-hot intergalactic medium (WHIM)
\cite[e.g.][]{2015arXiv150601031T}.

The effects of environment on galaxy evolution, the properties of
galaxies in clusters, and the formation of massive cluster central
galaxies have already been discussed in Section 5.  Here we focus on
additional science with DES clusters and existing auxiliary data.

\subsection{Cluster Evolution}

DES data (optical richness + ICL and lensing masses) coupled with X-ray and/or SZ samples of the hot gas phase will enable the study of the full baryonic content of massive haloes. The combined data will enable the study of the joint scalings of stellar mass, gas mass, gas temperature, X-ray luminosity, etc. with halo mass and redshift. In particular, the data should allow the covariance of these properties at fixed halo mass to be determined with significantly higher precision \citep{Evrard:2014}. In cosmological hydrodynamic simulations, the statistical relationships among stellar and hot phase properties are sensitive to the details of the ``sub-grid'' astrophysical models employed for star and supermassive black hole formation and their attendant feedback channels. Empirical constraints from DES studies will thereby lead to improved sub-grid prescriptions in simulations.

DES will also enable the study of cluster properties as a function of formation epoch and dynamical state, and will identify potentially early forming systems such as fossil groups \citep[e.g.][]{2003MNRAS.343..627J}.  Cluster formation epoch is thought to be an important additional parameter affecting the observed properties of clusters, such as richness, and contributing to the scatter in cluster scaling relations \citep{2015arXiv150302637M}.  Indicators for cluster formation epoch that can be probed by DES include the magnitude gap between the brightest and next brightest cluster galaxies, the luminosity of the brightest galaxy, and the offset between the brightest galaxy and the light-weighted mean galaxy position \citep{2014MNRAS.442.1578R}.  These measures can be used to correlate cluster richness and multiwavelength properties to cluster formation time.

In addition, DES will allow studies of the stellar mass fraction in clusters. This has already been done with early DES data in \citet{Palmese2016}, where they derived the stellar mass profile for the cluster RXC J2248.7--4431, and compared it with the total mass derived from  DES weak lensing analysis, resulting in a stellar mass fraction of $\sim 0.7$\% within  $r_{200c}\approx 3$ Mpc. 
This and similar techniques  will be applited to $\sim 10^5$ DES clusters to provide important information about galaxy evolution.

\subsection{AGN Evolution and Feedback}

In addition to the evolution of galaxies and their star formation histories as a function of environment, DES will also enable studies of AGN activity and studies of correlations between AGN activity and environment, galaxy and host cluster properties. Many observational lines of evidence point to a strong connection between supermassive black hole growth, star formation, and galaxy growth, including a similar evolution in AGN activity and star formation out to $z \sim2$ \citep[e.g.][]{Boyle1998,Franceschini1999,Merloni2004,Silverman2008}
 and the correlation between black hole mass and the velocity dispersion of the host galaxy's central bulge \citep[e.g.][]{Ferrarese2000,Gebhardt2000,Tremaine2002}.
AGN feedback is critical in balancing cooling and regulating star formation in both cluster cores and in galaxies;
 it is also clear that some form of non-gravitational heating is necessary out to large radii in clusters to explain the observed deviation of cluster scaling relations (such as the relation between the X-ray luminosity and the temperature of the intracluster medium) from the expected self-similar form \citep[e.g.][]{Arnaud1999,Markevitch1998}.
DES will enable studies of the connection of AGN activity to cluster and galaxy evolution with a large sample over a large redshift baseline, particularly in conjunction with existing X-ray and radio data.  Importantly, DES provides the ability to connect AGN activity to both galaxy environment and star formation.

The population of AGN in clusters (both in central and satellite
galaxies) compared to the field offers valuable information on AGN
fueling and the connection of AGN to galaxy evolution.  Previous
studies indicate a very strong evolution in the fraction of cluster
galaxies hosting X-ray detected AGN as a function of redshift with the
cluster AGN fraction increasing by a factor of 30 between $z\sim0.2$
and $z>1$ \citep[e.g.][]{2009ApJ...701...66M, 2013ApJ...768....1M}.
This evolution is stronger than the evolution in the field AGN
fraction so that at high-redshift the cluster and field AGN
fractions become comparable whereas at low redshift AGN are more common
in field galaxies \citep{2013ApJ...768....1M}.  This rapid evolution
in cluster AGN activity parallels a similar evolution in the fraction
of star forming galaxies in clusters over the same epoch \citep[e.g.][]{Haines2009}.
 In addition, there are some indications that groups and
smaller mass clusters at low redshifts host a higher AGN fraction than
massive clusters \citep{2009ApJ...707.1691A}.  These previous studies
are based on relatively small cluster and AGN samples.  Larger samples
of cluster AGN are already available in the relatively small area of
DES SV with overlapping Chandra archival data, and
preliminary results indicate a similar strong redshift evolution in
the cluster AGN fraction (Bufanda et al. in preparation).  The DES
cluster sample will also provide a much larger cluster mass range over
which to study the correlation of AGN activity to host halo mass.

\subsection{Cluster Sub-populations}

DES will discover new
clusters suitable for many diverse science programmes and will have the
statistics to identify samples of relatively rare sources.  DES
will, therefore, provide follow-up targets for a wide variety of
programmes including:

\textbf{Merging Clusters:} DES cluster samples combined with DES weak
lensing and/or existing X-ray and SZ data will allow the
identification of new bullet cluster-like systems suitable for tests of dark
matter self-interaction \cite[e.g.][]{bullet,2015Sci...347.1462H}.
The strong correlation of diffuse synchrotron radio emission with
cluster merging activity
\cite[e.g.][]{2015arXiv150603209C,2013ApJ...777..141C} also shows that
energetic cluster mergers are effective particle accelerators; however the
mechanisms responsible, whether shock acceleration, turbulent
re-acceleration, or secondary hadronic production are still poorly
understood.  Ongoing and upcoming surveys with the Low Frequency Array
(LOFAR) \citep{2013A&A...556A...2V} and Australian Square Kilometer
Array Pathfinder (ASKAP) \citep{2011PASA...28..215N} are expected to
allow the first large population studies of diffuse cluster radio
sources; the properties, particularly in terms of dynamical state, that are correlated with whether DES clusters host large-scale radio emission or not, will help illuminate the mechanisms responsible for particle
acceleration in clusters and the lifetime of diffuse radio emission.

\textbf{Relaxed Clusters:} The identification of the most relaxed
clusters, particularly at high redshifts, is important both for
cosmological studies as well as for understanding the balance of heating
and cooling in the ICM.  The fraction of cluster mass in baryons gives
a standard ruler by which the dark energy density and other
cosmological parameters can be constrained; this method requires
identifying relaxed clusters (which are relatively free of biases due
to asymmetry and lack of hydrostatic equilibrium) over a large
redshift baseline for deep X-ray follow-up
\cite[e.g.][]{2008MNRAS.383..879A}.  As highlighted above, the history
of cooling and AGN feedback in cluster cores shape the evolution of
the ICM.  Current evidence points to a lack of strong cool cores at
higher redshifts \citep[e.g.][]{2013ApJ...774...23M,
  2014ApJ...794...67M}, but larger samples and further study are
needed to confirm this trend.  Strong cool-core clusters such as the
Phoenix cluster \citep[SPT-CL J2344-4243;][]{2012Natur.488..349M} stand
out in the DES data as rich clusters with blue, star-forming central
galaxies.  The DES data will also allow studies of the central galaxy
properties of cool core clusters identified in X-ray or SZ.

\textbf{High-Redshift Clusters:} 
In combination with existing infrared surveys such as Spitzer Infrared Array Camera (IRAC) and WISE within the DES survey area, we will also be able to detect very high-redshift clusters in the range $z=1.0-1.5$ and possibly beyond.  The very deep data which DES is acquiring in the supernova fields will also provide additional leverage for high-redshift cluster science.  This sample will allow the study of clusters near the epoch of their formation. Little is known about clusters at redshifts much higher than $z\sim1$ as only small samples of very high-redshift clusters have been identified thus far, and the discovery of new high-redshift clusters by DES will have significant impact.  These clusters will allow us to extend many of the above mentioned studies to higher redshifts.  For example, while we observe significant evolution in the properties of cluster galaxies and the ICM from the present epoch out to $z\sim1$, many of the bulk cluster properties are already in place.  To understand questions of ICM heating, the establishment of the red sequence, and BCG formation it is advantageous to push our observations closer to cluster formation.

\section{Strong Gravitational Lensing}


DES expects to discover many new strongly lensed galaxies and quasars. A simple extrapolation from the CFHTLS-Strong Lensing Legacy Survey (using the 54 systems with rank 3 and above from \citealt{more12} and adjusting for the different survey depths) predicts that the wide survey will contain of order 1000 lenses with $18'' > \theta_{E} > 2''$ (where $\theta_{E}$ is the Einstein radius) of which about 200 would contain giant arcs (with arc length-to-width ratio $> 8$). A more rigorous prediction for galaxy-scale lenses \citep{collett15} suggests that we might discover of order 800 lenses with $\theta_{E}$ centred at about $1.5''$ using difference imaging between the $g$-band and the appropriately scaled $i$-band, as described in \citet{gavazzi14}. In addition we also expect to find a sample of $\sim120$ lensed quasars brighter than $i=21$ \citep{oguri10}, including twenty 
quadruples
with high information content
(we note that by  “brighter than $i=21$”  they define their limit in terms of the fainter component of pairs and third brightest components for quads).

\begin{figure*}
\center
	\includegraphics[width=12cm]{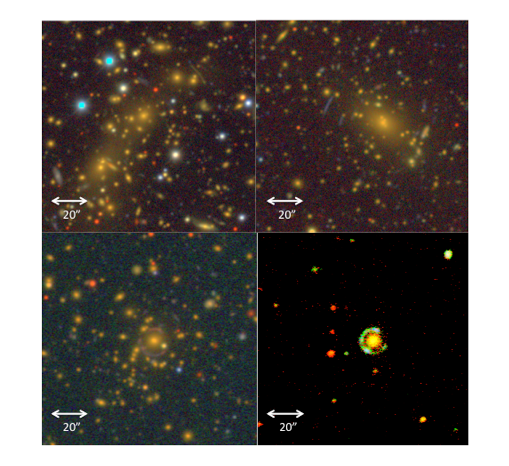}
    \caption{Four examples of known strong lensing systems from DES Science Verification and Year 1 Data.}
    \label{fig:lens_montage}
\end{figure*}

We are adopting a multi-pronged approach to identifying strong lenses in DES. To select lensed galaxies we will use a combination of automated arc-finders \citep[e.g.][]{gavazzi14,more12}, catalogue searches \citep[e.g.][]{kubik07,belokurov09,kubo10}, and crowd-sourced citizen-science efforts \citep{marshall15}. All of these have in common at least some amount of visual scanning as the final step in the search process. Examples of some known strong lenses in the DES footprint are shown in Figure~\ref{fig:lens_montage}. The lensed quasar search will use colour and morphology information at the catalogue level as well as the variable nature of quasars.  We have no $u-$band observations in DES, so to compensate we will need to use infrared bands, e.g. WISE. We have already carried out spectroscopic follow-up of some lens candidates from the DES SV data and have confirmed a number of systems \citep{Nord2015b}. In addition we have confirmed the first lensed quasars in DES, identified using the colour and morphology information \citep{agnello15}.

We will be able to use the sample of strongly lensed systems to study dark matter mass profiles at scales from individual galaxies all the way to massive clusters. Cluster-scale lenses are particularly useful because they allow us to study the effects of strong lensing in the core of the cluster and weak lensing in the outer regions. Strong lensing provides constraints on the mass contained within the Einstein radius of the arcs whereas weak lensing provides information on the mass profiles in the outer reaches of the cluster.  Combining the two measurements allows us to make tighter constraints on the mass and the concentration of an Navarro-Frenk -White (NFW, \citep{NFW}) model of the cluster mass density profile over a wider range of radii than would be possible with either method alone \citep{natarajan98,natarajan02,bradac06,bradac08a,bradac08b,diego07,limousin07,hicks07,deb08,merten09,oguri09}. In addition, if one has spectroscopic redshifts for the member galaxies one can determine the cluster velocity dispersion, assuming the cluster is virialized, and hence obtain an independent estimate for the cluster mass \citep{becker07}.  These different methods, strong plus weak lensing and cluster velocity dispersion, provide independent estimates of the cluster mass and can then be combined to obtain improved constraints on the mass and concentration \citep[e.g.][]{buckleygeer11}. Galaxy-scale lenses with multiple source planes \citep{gavazzi08} also provide excellent constraints on the dark matter distribution in galaxies \citep{sonnenfeld12}, which is a sensitive test of galaxy formation models and baryonic feedback processes \citep{johansson12}.

We will have a large sample of sources at varying redshifts which will be very valuable for studies of galaxy evolution \citep[e.g.][]{saintonge13,genzel14,bayliss14,rhoads14}. For example \citet{collett15} predicts that the source redshift distribution for galaxy-scale lenses peaks at about $z_{s} = 2.0$.  Strongly lensed galaxies are particularly useful for the study of high-redshift galaxies due to the magnification of the galaxy flux. Therefore, lensed galaxies are prime candidates for detailed studies, since they can be investigated using only a fraction of the telescope time that would be needed to study comparably distant but unmagnified galaxies. Moreover, since lensing increases the apparent angular size of the source, it enables the study of source galaxy substructure with greater effective spatial resolution than for an un-lensed source of the same physical size at comparable distance. The excellent red-sensitivity of the DECam CCDs, along with the $grizY$ filter set, provides sensitivity to high-redshift Lyman Break Galaxies (e.g., $g$-band dropouts at $z\sim 4$ or $r$-band dropouts at $z\sim 5$).

One of the primary motivations for the lensed quasar search is cosmology \citep[e.g.][]{suyu13}. However considerable additional science can be carried out using these systems. For example, stars in the lens cause microlensing effects on the multiple quasar images, which can be used to examine the stellar contents of the lens galaxy \citep[e.g.][]{oguri14, schechter14,jimenez15}. The same microlensing can also provide constraints on the inner structure of the lensed quasar, such as the size of the accretion disk and the thermal profile \citep[e.g.][]{poindexter08} as well as the geometry of the broad line region \citep[e.g.][]{sluse11}. In addition, milli-lensing caused by flux ratio anomalies probes the presence of substructure in the lens \citep[e.g.][]{nierenberg14}. Finally, using high resolution imaging of the lens allows the reconstruction of the source and can provide a direct view of quasar-host coevolution up to $z\sim2$ \citep[e.g.][]{Peng2006,rusu15}.

\section{Quasars}



The DES photometric data will contain millions of new quasars (QSOs), 
and in particular new samples at $z>6$ and $z>7$ that can be used to study the 
reionisation history of the Universe through measurements of the transparency 
of the intergalactic medium (IGM), the evolution of supermassive black hole 
masses, the chemical evolution of the IGM from the statistics of metal-line 
absorption systems, their space density evolution, and many other topics. In 
addition, the regular monitoring cadence employed for the supernova survey 
will include variability data that can be used to estimate supermassive black hole 
masses via the reverberation mapping technique, as well as to select quasar candidates.
Figure 13 shows a DES colour-colour diagram, as well as known quasars overlapping the DES SV observations.

\subsection{Expected Numbers of Quasars}

We have constructed mock quasar catalogues for DES using the quasar
spectral energy distribution models of \citet{Maddox:08, Maddox:12}
and the quasar luminosity functions from \citet{Ross:12} at $z<5$,
\citet{McGreer:13} at $5<z<6$ and \citet{Willott:10} at $z>6$. Based
on these luminosity functions (or extrapolations at high redshifts
assuming pure density evolution), we can calculate the total number of
quasars over various redshift ranges that can be expected in the DES
volume. 
Illustrative sample sizes 
of photometrically selected quasars 
over the DES wide survey area of 5000 sq deg and down to the typical 10-sigma point source depths  include
$ ~670,000$ with $ i_{AB} <24.0$ at $ 2 < z < 4$, 
$~33,000$ with $z_{AB} < 23.4$ at $4< z< 5$, 
$~370$ with $z_{AB}  < 23.4$ at $6 < z < 7$, 
and 
$~5$ with  $Y_{AB}<21.7$  at $z > 7$.

In addition, the DES supernova fields covering $\sim$27 sq deg are
imaged to much fainter magnitude limits. Although DES lacks $u$-band
coverage, the $grizY$ filters can effectively be used for the
selection of faint $z>4$ quasars (which appear as $g$-band drop-outs)
and $z>5$ quasars (which appear as $r$-band drop-outs). Down to a
5$\sigma$ depth of $z<26.5$, we expect $\sim$840 quasars at $4<z<5$ and $\sim$390 quasars at $5<z<6$ in the DES supernova
fields. At these magnitude limits, the faint quasar population begins
to overlap the bright end of the Lyman Break Galaxy (LBG) population,
thus potentially enabling constraints on the relative space densities
of the brightest star-forming galaxies and the faintest active
galactic nuclei (AGN) observable at high redshifts.

\begin{figure}
\begin{center}
\includegraphics[scale=0.5]{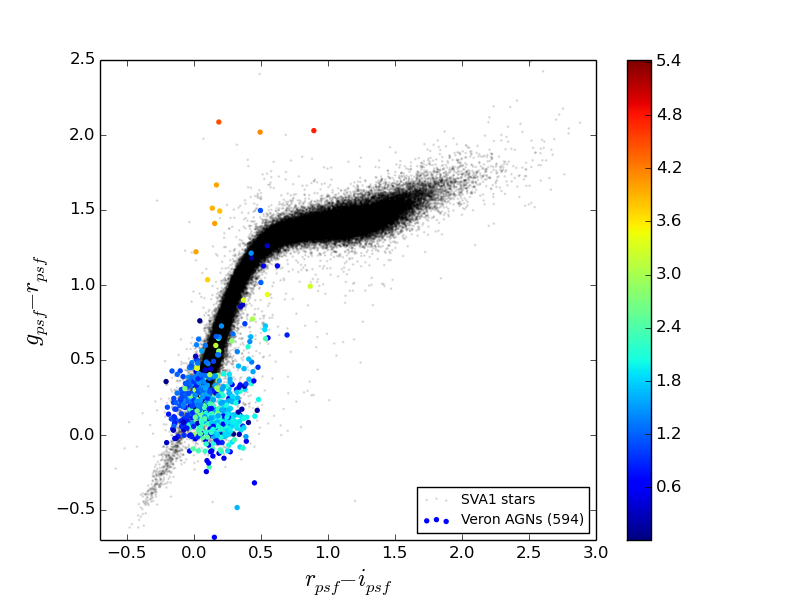}
\caption{
DES $gri$ colour-colour diagram showing the stellar locus as well as known quasars overlapping the DES Science Verification observations. 
The quasars have been colour-coded according to redshift. While the low redshift quasars overlap the stellar locus, the $z>4$ quasars clearly appear as very red $(g-r)$ sources.}
\label{gri}
\end{center}
\end{figure}

\subsection{Quasars at the Epoch of Reionization}

The very deep $z-$band data obtained as part of DES, combined with $Y-$band 
observations, make the DES data particularly well-suited for the discovery 
of new quasars at redshifts of six and above. At these redshifts, quasars may be 
selected as $i-$band dropouts, or even $z-$band dropouts.  The source of 
confusion for these dropout samples are very red Galactic stars and substellar 
objects, namely L and T dwarfs \citep{Fan:01, Fan:06}, which have 
substantially higher surface densities at these magnitude limits. Quasars may 
be separated from stellar objects in colour-space with the addition of a 
redder filter, as the intrinsic SED of quasars is generally much bluer than 
these Galactic objects. For $z \sim 6$ quasars, $z-Y$ colours are an effective 
discriminant, while for the $z \sim 7$ quasars that will be $z-$band dropouts
\citep{Mortlock:11}, 
$J-$band observations from VHS and VIKING will be used.

\citet{Reed2015} illustrated the promise of quasar discovery with DES
with the discovery of a quasar at $z > 6$. The quasar was the
brightest of 43 candidates and was identified for follow-up work
solely on the DES $i-z$ and $z-Y$ colours. Spectroscopy of the quasar,
DES J0454-4448, showed that it has an emission redshift of $z=6.10 \pm
0.03$ and a HI near zone size of $4.6 \pm 1.7$ Mpc.  The presence of
one $z>6$ quasar in the relatively small area probed by the DES
SV data are consistent with the luminosity function
expected at $z \sim 6$.

Quasars at redshifts below six show a substantial amount of
Lyman-$\alpha$ absorption due to intervening neutral hydrogen gas at
lower redshift, yet the transmission of some quasar continuum below
the break indicates that the IGM is mostly ionized. In contrast, quasars at
redshift six and above exhibit complete Gunn-Peterson absorption
troughs \citep{Gunn1965} that indicate that the optical depth is much
greater than one and imply a significant change in the neutral
fraction \citep{Fan2002, Fan:06, Meiksin2009}. However the present
data are unable to differentiate between a neutral fraction of 1\% and
100\%, or whether there are comparable volume fractions at these two
extremes. In addition to the evidence for a global increase in the
neutral fraction, some sight lines show regions with substantial
transmitted flux above redshift six \citep{White2003}. This variable
optical depth is expected from models of reionisation, which predict
substantial variations in the neutral fraction when the Str\"omgren
spheres are growing, but do not yet overlap
\citep[e.g.][]{Furlanetto2004, Furlanetto2009}. The new, bright quasars
discovered with DES will be valuable for improved determinations of the patchiness
of the ionisation fraction above redshift six, the space
density of the sources that produce reionisation, whether reionsation
begins in overdense or underdense regions, and the physical state of
the IGM.

\subsection{Space Density Evolution}
Another interesting result to arise from the study of quasars at 
$z>2.9$ is that their correlation lengths are as large as those for massive 
clusters of galaxies in the local universe, and well above those measured 
for local quasars \citep{Shen2007}. As the quasar space density also declines 
dramatically at $z>2.9$ \citep{Warren1994, Schmidt1995, Richards2006}, the combination of 
very low space density and very large correlation length provide 
very strong constraints on the fraction of the time galaxies appear as quasars (the duty 
cycle) and the mass accretion history of 
their supermassive black holes \citep{Cole1989, Haiman2001, Martini2001}. These 
measurements also place a strong constraint on the maximum amount of scatter 
between quasar luminosity and halo mass \citep{White2008}. \citet{Shankar2010}
found that a model with the duty cycle equal to 0.2, 0.5, and 0.9 at 
redshift 3.1, 4.5, and 6.0 respectively provides the best match to the clustering and 
space density data. In the context of their model, part of the reason 
that the quasar space density drops precipitously between redshift three and 
six is because sufficiently massive haloes are progressively more rare, yet 
the rate of this decline is partially compensated by the increase in the 
duty cycle. As the duty cycle cannot exceed unity, the decline in the quasar 
space density above redshift six is predicted to be yet more precipitous 
because it will be forced to track more exactly the evolution of the 
most massive haloes. This space density decline above redshift six is approximately a factor of 
three faster than a simple extrapolation of the space 
density evolution from three to six. The discovery of a substantial number 
of quasars at redshift seven, and the first measurement of the quasar space 
density at this redshift, will provide a powerful test of this model 
prediction and thereby shed light on the relationship between massive haloes and quasars 
at the highest redshifts. 

\begin{figure*}

\begin{center}
\begin{tabular}{c}
\includegraphics[scale=0.5]{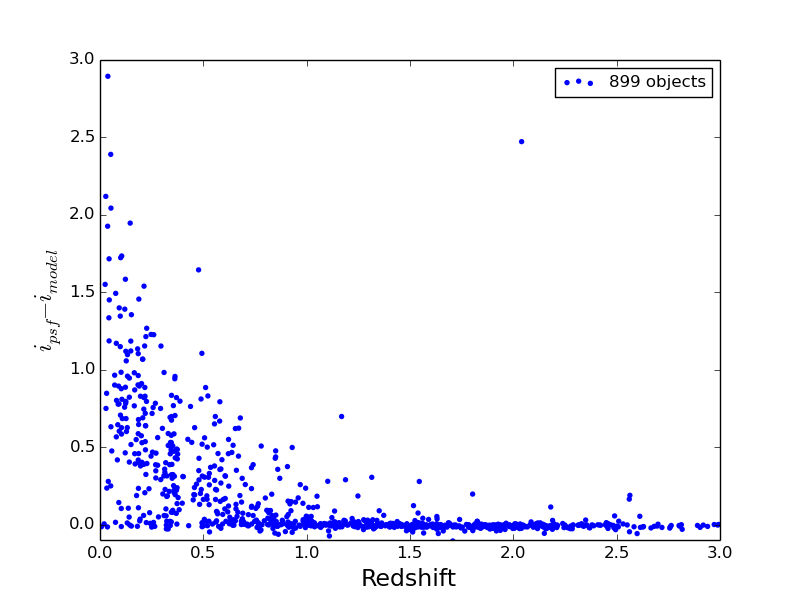} \\
\end{tabular}
\begin{tabular}{cc}
\includegraphics[scale=0.3]{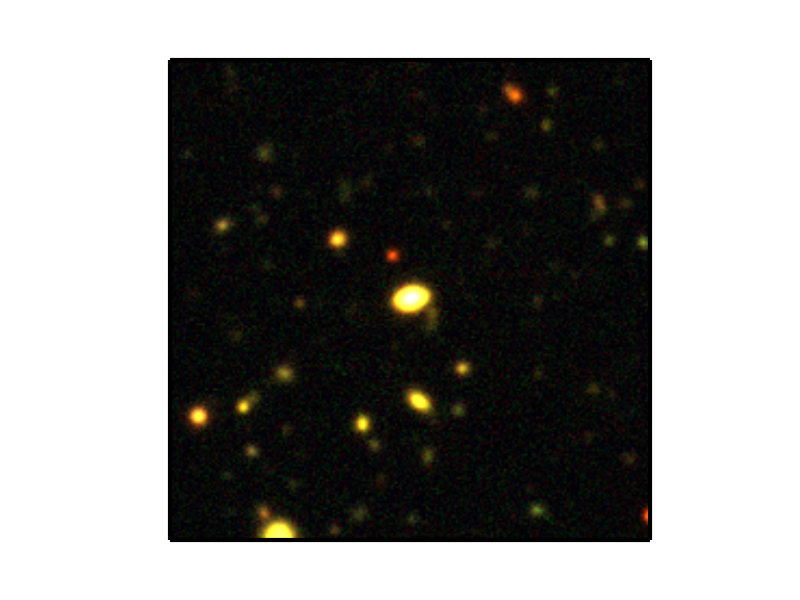} & \includegraphics[scale=0.3]{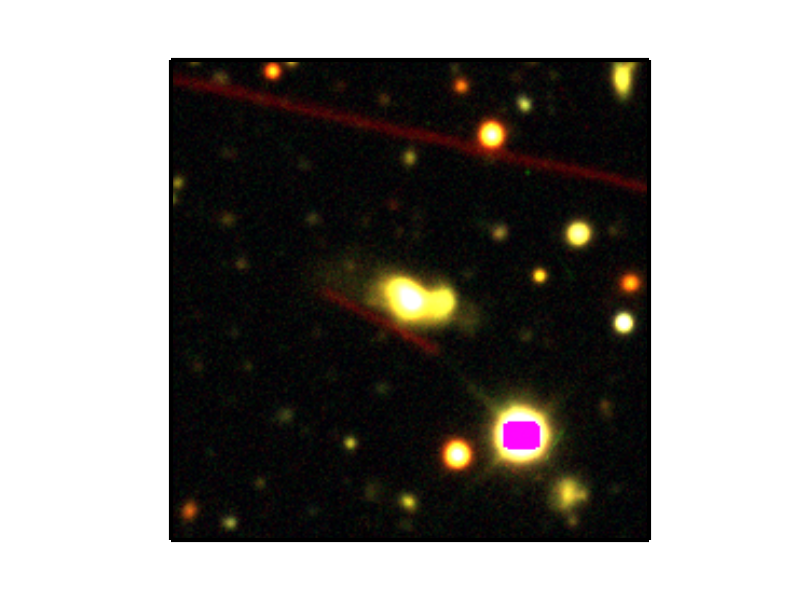} \\
\end{tabular}
\caption{
Extended objects can be identified by comparing the psf magnitude to the model magnitude.  
{\it Top:} The figure shows that many quasars in our sample are resolved (appear to be extended objects) even out to $z\sim1.0$.
{\it Bottom:} DES $gri$ colour-composite images of two known quasars at $z=0.85$ (left) and $z=1.17$ (right). Both quasars have extended morphologies in the DES data demonstrating that low surface-brightness host galaxy features are clearly visible for quasars out to $z\sim1$.}
\label{fig:qsohost}
\end{center}
\end{figure*}

\subsection{Black-Hole Mass Calibration via Reverberation Mapping}
A fundamental challenge for studies of the evolution of supermassive
black holes is accurate estimates of their masses. Dynamical
measurements of supermassive black hole masses are possible for
quiescent galaxies at distances up to about 100 Mpc for the largest
telescopes working near the diffraction limit \citep{McConnell2012}. However similar measurements are not feasible for objects at larger
distances, and especially not cosmological distances ($z>0.1)$, and
there are thus no direct, dynamical measurements of how supermassive
black holes grow over cosmic time. The best objects to use in the study of the
evolution of supermassive black hole mass are broad-line AGN, which
can be observed to the highest redshifts. Reverberation mapping
\citep{Blandford1982, Peterson1993} provides the most direct
measurement of black hole mass for AGN. This technique employs
measurements of the velocity width of a broad emission line and the
time lag between continuum and line luminosity variations. The time
lag provides the size scale of the broad line region, and combined
with the velocity width of the broad line, provides a measurement of
the enclosed (black hole) mass. While there are inherent uncertainties
due to the geometry and dynamics of the broad-line region
\citep{Horne2004, Collin2006, Pancoast2014}, these measurements agree well with mass
estimates based on stellar kinematics \citep{Ferrarese2001, Onken2014}
and have been used to develop empirical relationships between
continuum luminosity (which is used as a proxy for broad-line region
size) and line width that have been employed to estimate the evolution
of supermassive black holes over cosmic time \citep{Kaspi2002,
  McLure2004, Vestergaard2006, Kollmeier2006, Vestergaard2009}.  The
key uncertainty in this extrapolation to high redshifts is that the
local AGN studied via reverberation mapping have almost exclusively
been low-luminosity objects \citep{Dasyra2007, Delmoro2008, Bentz2013, Bentz2015}, while
studies of black hole evolution employ quasars that are an order of
magnitude more luminous. This is both because of the scarcity of truly luminous AGN in the local universe and because the variable timescales for such AGN are substantially longer and consequently present a greater practical challenge. The 27 sq deg DES SN survey is well
suited to measure black hole masses for higher-luminosity and higher
redshifts due to the regular cadence of photometric observations for
five years, combined with approximately monthly spectroscopic
follow-up observations as part of the OzDES survey \citep{Yuan2015} on the Anglo-Australian Telescope (AAT). There are
several thousand quasars in this area with $0.1 < z < 4$ and $i < 21$
mag and the DES observations of the SN fields provide regular
monitoring of them for the substantial continuum variations that
provide a strong reverberation signal. At the same time, regular
spectroscopic follow-up observations of $\sim 800$ of these quasars
obtained as part of the OzDES survey will be used to measure the
corresponding emission-line flux variation and thus the size of the
broad-line region. Based on detailed simulations of this survey by
\citet{King2015}, we expect to recover lags for 35-45\% of these
quasars. Reverberation-based black hole masses for even a small
fraction of these quasars would substantially increase the present
sample of approximately $\sim 70$ AGN with such data \citep{Bentz2010,
  Grier2012, Barth2015, Bentz2015}, as well as populate the largely unconstrained
parameter space at high luminosities.

\subsection{Quasar Host Galaxies}

One of the main questions in AGN research concerns how supermassive
black-holes are fueled. For luminous quasars, the necessary fuel
supply is substantial (a few M$_\odot$yr$^{-1}$) and estimates of the
typical quasar lifetime suggest that this accretion rate is maintained
for $\sim$10$^{7-8}$ years \citep{Martini2004}. Gas-rich galaxy
mergers appear to be the most viable fueling mechanism
\citep[e.g.][]{Dimateo2005, Hopkins2008} in galaxy formation
models. However, observational evidence for quasars being hosted in
merging galaxies has proved elusive, particularly at high
redshifts. This is largely because the bright quasar emission makes it
challenging to classify the host galaxy morphology in all cases, as
well as because of the absence of a comparison sample to quantify the incidence
of mergers. Data from DES can help to address both of these
issues.
DES
data will allow us to resolve low surface brightness features far from
the quasar nuclear flux. A census of merging systems in non-quasar
hosts over the same redshift range can also be extracted from the DES
data and used as a control sample. This control sample is important to
quantify the fraction of all mergers that are detectable at the larger
radii probed by quasar hosts. Figure \ref{fig:qsohost} shows the
difference between the PSF and model magnitude in the $i$-band for a
sample of known luminous quasars overlapping DES SV data from
\citet{Veron1993}. As the PSF magnitude only encompasses flux in the
central nuclear region whereas the model magnitude is a better measure
of the total flux, unresolved point sources will have a small
difference between their PSF and model magnitudes whereas extended
sources will have a much larger difference. The top panel of Figure
\ref{fig:qsohost} shows that at high redshifts, luminous quasars are
unresolved in DES data as expected. However, extended morphologies are
seen for a significant number of quasar host galaxies out to
$z\sim1$. Two example DES images for $z\sim1$ quasars are also shown
in Figure \ref{fig:qsohost} and demonstrate that DES will be uniquely
sensitive to quasar host galaxies at these redshifts.

Host galaxy studies will be even more viable in dust-reddened quasars
\citep[e.g.][]{Banerji2012, Banerji2013} where there is significant
extinction towards the quasar line of sight. As these dusty quasars
are thought to correspond to a phase of transition between starburst
galaxies and UV-luminous quasars, a large fraction of them are
expected to be hosted in merging systems. At low redshifts, there is
evidence that dust-reddened quasars do indeed reside in mergers
\citep{Urrutia2008} and DES will enable an extension of these studies
to much higher redshifts.

\section{Supernovae and Transients}

The DES Supernova programme (DES-SN) was designed to discover and
photometrically follow thousands of Type Ia Supernova
(SNe~Ia) for cosmology \citep{Bernstein2012}.  The DES-SN programme
surveys 27~sq~deg over 10 DECam fields (two `deep,' eight `shallow')
chosen to overlap with well-observed fields: the ELAIS-S, XMM-LSS,
CDF-S, and `Stripe82' with the latter three also corresponding to  PS1
Medium-Deep Fields. 
The survey has already discovered
$> 1000$ SN~Ia candidates, with hundreds of host-galaxy redshifts
obtained via the OzDES project \citep[see][]{Yuan2015} and nearly 100 supernovae spectroscopically classified using other 4 -- 10-m
class telescopes.

While running the SN survey, DES will naturally discover thousands of
other astrophysical transients.  The largest fraction should be
core-collapse SNe.  However, the survey will also discover many
``exotic'' transients (those not falling into the normal SN classes of
II, Ia, Ib, or Ic).  In particular, DES-SN has already discovered
several superluminous SNe \citep[SLSNe;][]{Papadopoulos2015, Smith2015}
and one tidal disruption event \citep[TDE;][]{Foley2015}.  Here we
briefly outline the potential science that can be achieved with these
discoveries.

As of Y3 we are now making public instantaneously all our bright ($r<20$) transients at https://portal.nersc.gov/des-sn/ 
so that they can be followed up by other groups.

\subsection{Core-Collapse SNe}

Core-collapse SNe, those SNe with massive-star progenitors that undergo
a core collapse, represent roughly 75\% of all stellar explosions in a
given volume \citep{Li11:rate2}.  However, core-collapse SNe tend to
be fainter than SNe~Ia, and thus only represent about 20\% of all
discovered SNe in a magnitude-limited survey \citep{Li11:rate2}.
Nonetheless, DES-SN should discover about 500 core-collapse SNe.

There are currently several open questions about core-collapse SNe,
ranging from their precise progenitors -- especially for different
classes \citep[e.g.,][]{Smartt09:review}, how they explode
\citep[see][for a review]{Janka07}, and their connection to gamma-ray
bursts \citep[e.g.,][]{Woosley06}.  Furthermore, core-collapse SNe can
be used to track the star-formation rate history of the Universe
and possibly for constraining cosmological parameters
\citep[e.g.,][]{Nugent06}.

DES-SN has already spectroscopically classified
many ``typical''
core-collapse SNe.  DES-SN should be able to detect most core-collapse
SNe out to $z \approx 0.2$.  With the full photometric sample, we plan
to address the above topics.

\subsection{Superluminous SNe}

So far DES has published results for two 
superluminous supernovae (SLSNe).
There are others confirmed spectroscpically
\citep[e.g. DES15E2mif,][]{Pan2015} 
and many more candidates.

The first spectroscopically confirmed SLSNe
discovered by DES (DES13S2cmm) is presented in detail in
\citet{Papadopoulos2015}.  The event was discovered early in the first
year of DES operations.  Spectroscopy with the ESO Very Large
Telescope (VLT) showed that DES13S2cmm was spectroscopically similar
to SLSN-I (meaning it lacks hydrogen; see \citet{Gal-Yam2012} for
details of SLSN classification) at $z = 0.663 \pm 0.001$ (with the
redshift from host-galaxy emission lines).  At this redshift, the SN
had a peak $U$-band rest-frame absolute magnitude, $M_{U, {\rm peak}} =
  -21.05^{+0.10}_{-0.09}$~mag, consistent with other SLSNe.

Like some other SLSNe \citep[e.g.,][]{Neill11, Lunnan14:host} ,
DES13S2cmm is located in a faint, low metallicity (sub-solar), low
stellar-mass host galaxy with $\log(M/M_{\sun}) = 9.3 \pm 0.3$
\citep[see][]{Lunnan2015}.  The bolometric light curve of DES13S2cmm
is consistent with being powered by the spin down of a magnetar, see 
\citet{Kasen10:mag} and \citet{Woosley10},
a model which can fit the basic observables of several SLSNe-I \citep[e.g.,][]{Nicholl2014}.


The second published DES SLSN is DES14X3taz \citep{Smith2015}
 which is also hydrogen-poor (SLSN-I) at a redshift of $z=0.608$. The light curve of DES14X3taz is double-peaked, with the first peak being relatively blue, lasting 15 rest-frame days, with a peak luminosity of 30\% of the main second peak which resembles other SLSN-I in DES
\citep{Papadopoulos2015}  and the literature. The discovery of double-peaked SLSN light curves provides an important insight into their possible progenitors, suggesting an initial very hot event (first peak) that cools rapidly followed by the re-heating of the very extended stellar envelope (hundreds of solar radii) by a central engine such as a magnetar. Such double-peaked SLSNe could be common as recently suggested by \citet{Smith2015} 
  and \citet{Nicholl2015}.
  Based on current rate estimates \citep{Quimby2013}
  , DES should detect between 15 to 20 SLSNe over 5 years \citep[see][]{Scovacricchi2015}.

\begin{figure}
\begin{center}
\includegraphics[scale=0.40]{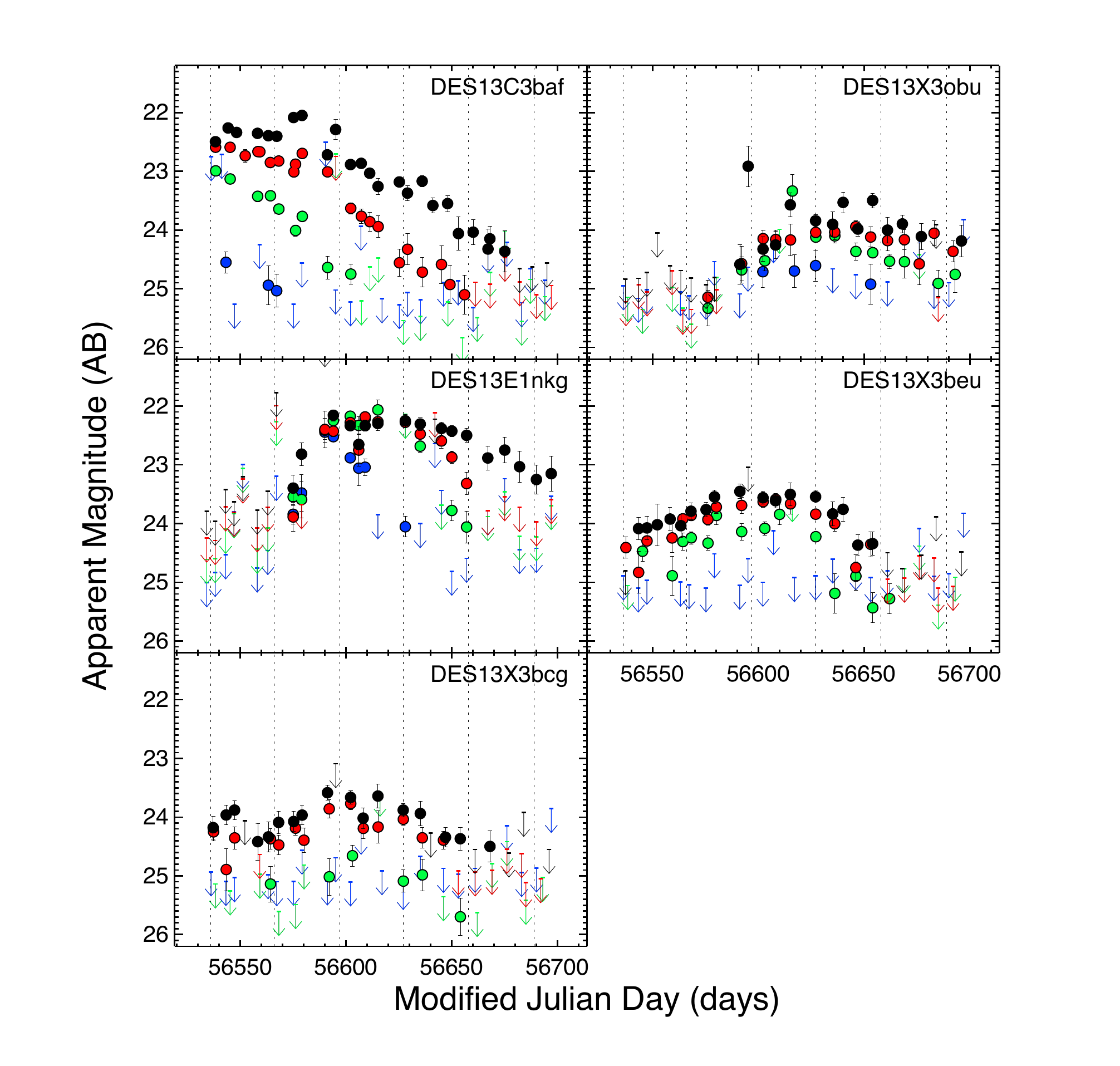}
\caption{Examples of five hostless SLSNe candidates selected from Y1 of DES. These are in addition to DES13S2cmm \citep{Papadopoulos2015} which was spectroscopically confirmed. The long multi-colour light curves of these candidates are consistent with other SLSNe and are not AGNs due to the lack of a visible host galaxy. The colours are blue (g-band), green (r-band), red (i-band) and black (z-band). Upper limits are shown as arrows.}
\label{THESIS_MAG_LCs}
\end{center}
\end{figure}

 In Y1, we started a programme for selecting SLSNe
candidates from the pool of all DES-SN detected transients using the
following criteria: (i) At least one month of multicolour data
(typically five to six detections with S/N $> 3.5$ in each of $griz$); (ii)
a low 
probability of fit to any standard type Ia, II, or IIb/c light curve
\citep{Sako2014}; (iii) SN located at least one DECam pixel from the
centre of its host galaxy (if a host was detected); and (iv) a peak
observed magnitude no fainter than one magnitude below its host galaxy
(if detected).  This resulted in up to 50 possible candidates which
were then inspected visually to determine if their multicolour light
curves were consistent with other observed SLSNe, i.e., if the rise
time of the light curve was greater than 25 observer-frame days and,
provided a host photometric or spectroscopic redshift was available,
if the peak absolute magnitude was less than $-20.5$~mag.  
See Figure~\ref{THESIS_MAG_LCs} for examples of five hostless SLSNe candidates from Y1.

The search for SLSNe candidates continued into Y2 of the DES-SN
programme and, due to increased spectroscopic follow-up time, we have
spectroscopically confirmed many more SLSNe candidates.
New surveys such as ``Search Using DECam for Superluminous Supernovae''
(SUDSS), which supplement the DES-SN data should increase the yield of
SLSNe to nearly a hundred such events out to $z \approx 2$
\citep{Scovacricchi2015}. 
The size and quality of such a sample should
greatly improve our astrophysical knowledge of these luminous
explosions.

\subsection{Exotic Thermonuclear Transients}

SNe~Ia are thermonuclear explosions of C/O WDs.  There
are several other classes of explosions with similar origins.  Of the
classes which have been observed, there are the low-energy,
low-luminosity SNe~Iax \citep{Foley13:iax}, the Ca-rich SNe
\citep{Perets10:05e} which may come from stars ejected from their
galaxies \citep{Foley15:carich}, the low-velocity, low-temperature
SN~2002es-like SNe \citep{Ganeshalingam12}, the troublesome
SN~2006bt-like SNe \citep{Foley10:06bt}, and kilonovae \citep[which
are thought to have neutron-star progenitors;][]{Berger13:kilo,
  Tanvir13}.  There are also several proposed classes such as Ia SNe
\citep{Bildsten07}.

Exotic transients are not just an idle curiosity; they are the result
of the most extreme physical conditions imaginable.  For instance, the
above classes are the result of truly unexplored phenomena and
represent new endpoints of stellar evolution and the boundaries of
possible physical conditions of stellar systems.  Meanwhile, the
explosions probe interesting corners of physics such as $r$-process
element creation.  Since many exotic transients share some similar
properties to SNe~Ia, one may learn more about the physics of SNe~Ia
by observing their peculiar cousins.

Many of the transients in these classes are less luminous and faster
than SNe~Ia.  Moreover, all appear to have a lower volumetric rate
than SNe~Ia \citep[e.g.,][]{Foley13:iax}.  As a result, exotic
thermonuclear transients are generally detected at a rate of
$\lesssim$5\% that of SNe~Ia in magnitude-limited surveys
\citep{Li11:rate2}.  For the DES-SN survey, we therefore expect to
detect $\sim$ 100 exotic thermonuclear transients.  Because of their
low luminosity, nearly all should be at $z \lesssim 0.1$.  DES-SN has
not yet spectroscopically identified any such transient.

\subsection{Tidal Disruption Events}

If a star comes sufficiently close to a massive black hole (MBH), the
tidal forces will be strong enough to disrupt the star.  If this
occurs outside of the event horizon, it is possible that a luminous
transient will be associated with the disruption.  \citet{Rees88}
first described this process, showing that for stars on bound orbits,
that roughly half of the disrupted star would become unbound, while
the other half would return on bound orbits, eventually accreting onto
the MBH.  This accretion should power a luminous transient similar to,
but generally stronger than an AGN flare.

Tidal disruption events (TDEs) are useful for detecting quiescent
BHs, which is particularly important for determining the MBH
occupation fraction in low-mass galaxies.  Since the luminosity of a
TDE is related to the mass of the MBH, a large sample of well-observed
TDEs can be used to cleanly measure the MBH mass function.
Furthermore, the TDE rate, especially in recently merged galaxies, can
indicate the conditions of the nuclear star cluster and determine the
importance of stellar scattering for MBH mergers.

There have been several tidal disruption event (TDE) candidates
discovered in the optical \citep[e.g.,][]{VanVelzen11, Arcarvi14,
  Holoien14}, with the best-observed examples being discovered by PS1
\citep{Gezari12, Chornock14:tde} and ASAS-SN \citep{Holoien16}.  With
multi-colour light curves that cover the entire rise and decline, one
can begin to disentangle parameters of the disrupted star (such as
mass, composition, impact parameter, etc) and the MBH mass.  There
have been few TDEs with good optical coverage on the rise, and
additional data will be critical for understanding the physics of
TDEs.

To date, DES-SN has detected a single TDE candidate
\citep{Foley2015}.  Based on current TDE rates
\citep[e.g.,][]{vanVelzen14}, we expect 1--5 TDEs discovered in the
DES-SN survey.

\subsection{Gamma Ray Bursts}

We have initiated a programme to provide to the astronomical community in near-real time those data that are relevant to transients detected by other instruments. In its initial application, this programme entitled DESAlert \citep{Poci2015}  automatically receives VOEvent notices \citep{Seaman2011} from the Swift satellite \citep{Barthelmy2005} whenever its instruments detect a gamma-ray burst (GRB). In response, DESAlert finds all objects within the DES catalogs within a pre-determined search radius of the GRB position, and disseminates information about these objects in a subsequent VOEvent notice of its own. Of particular relevance to follow-up observations of these GRBs, DESAlert identifies potential standard calibration stars (facilitating immediate relative photometry), as well as potential host galaxies (including magnitude, photometric redshift, and morphological characteristics). The DESAlert service is expected to provide this information to the community for 1-2 dozen GRBs (all those occurring within the DES footprint) every year.

\section{Summary}


In this overview article we have illustrated the legacy prospect of DES, beyond its primary goal for cosmological studies.
DES is providing both a deep mapping of 5000 sq deg  of the sky, as well as a
time-domain survey over 27 sq deg.  Although optimized to measure and
characterise dark energy, it has already yielded both expected and
unexpected discoveries using initial SV data 
 and the Y1 \&Y2  data over most of the footprint (to about 40\% of the expected 5yr magnitude depth).

We have highlighted some of the
initial discoveries, from the solar system to the high redshift universe (see Table 1 for more details):
\begin{itemize}
\item{34 Trans-Neptunian Objects (TNO)};
\item{17 new dwarf satellites of the Milky Way, and mapping of the LMC and its stellar population};
\item{Two published superluminous supernovae, and more confirmed};
\item{One published quasar at redshift $z> 6$, and more confirmed};
\item{Two strongly lensed quasars.}
\end {itemize}

The forecast for the complete 5yr DES is very promising; for example, 
DES is likely to discover:
\begin{itemize}
\item{Many times more `hot' TNOs than any previous survey};
\item{More new dwarf satellites of the Milky Way, including even lower surface brightness objects than are currently known, as well as new stellar streams that probe the MW halo};
\item{Over 370 quasars at redshift $6 < z <  7 $ and $\sim$ five at $z >7$};
\item{roughly $1000$ strong lensing systems}.
\item {Possible detection of optical counterparts of gravitational wave events}.
\end {itemize}

In addition, the cosmological probes of the wide-field DES, i.e. galaxy
clustering, weak lensing and clusters, have already produced
interesting results on galaxy biasing as a function of redshift, and
on the profiles of galaxy versus dark matter in clusters.  We expect that
analysis of the full dataset will allow tight constraints on the
galaxy-halo connection in a broad range of environments and out to $z
\sim 1$, and will produce new insights into galaxy and quasar
evolution.

DES data are powerful not just on their own, but also in combination with other multi-wavelength data
 (e.g. VISTA, SPT) and
spectroscopy (e.g. OzDES, 2dFLens).  The discoveries already made by
DES using advanced statistical techniques also foretell the
great discoveries ahead with the next generation of deeper and wider
imaging surveys such as  LSST, Euclid and WFIRST, combined with spectroscopic surveys such as DESI.

\section*{Acknowledgments}

Funding for the DES Projects has been provided by the U.S. Department
of Energy, the U.S. National Science Foundation, the Ministry of
Science and Education of Spain, the Science and Technology Facilities
Council of the United Kingdom, the Higher Education Funding Council
for England, the National Center for Supercomputing Applications at
the University of Illinois at Urbana-Champaign, the Kavli Institute of
Cosmological Physics at the University of Chicago, the Center for
Cosmology and Astro-Particle Physics at the Ohio State University, the
Mitchell Institute for Fundamental Physics and Astronomy at Texas A\&M
University, Financiadora de Estudos e Projetos, Funda{\c c}{\~a}o
Carlos Chagas Filho de Amparo {\`a} Pesquisa do Estado do Rio de
Janeiro, Conselho Nacional de Desenvolvimento Cient{\'i}fico e
Tecnol{\'o}gico and the Minist{\'e}rio da Ci{\^e}ncia, Tecnologia e
Inova{\c c}{\~a}o, the Deutsche Forschungsgemeinschaft and the
Collaborating Institutions in the Dark Energy Survey.  The DES data
management system is supported by the National Science Foundation
under Grant Number AST-1138766.

The Collaborating Institutions are Argonne National Laboratory, the
University of California at Santa Cruz, the University of Cambridge,
Centro de Investigaciones En{\'e}rgeticas, Medioambientales y
Tecnol{\'o}gicas-Madrid, the University of Chicago, University College
London, the DES-Brazil Consortium, the University of Edinburgh, the
Eidgen{\"o}ssische Technische Hochschule (ETH) Z{\"u}rich, Fermi
National Accelerator Laboratory, the University of Illinois at
Urbana-Champaign, the Institut de Ci{\`e}ncies de l'Espai (IEEC/CSIC),
the Institut de F{\'i}sica d'Altes Energies, Lawrence Berkeley
National Laboratory, the Ludwig-Maximilians Universit{\"a}t
M{\"u}nchen and the associated Excellence Cluster Universe, the
University of Michigan, the National Optical Astronomy Observatory,
the University of Nottingham, The Ohio State University, the
University of Pennsylvania, the University of Portsmouth, SLAC
National Accelerator Laboratory, Stanford University, the University
of Sussex, and Texas A\&M University.

OL acknowledges support from a European Research Council Advanced Grant FP7/291329.
The DES participants from Spanish institutions are partially supported
by MINECO under grants AYA2012-39559, ESP2013-48274, FPA2013-47986,
and Centro de Excelencia Severo Ochoa SEV-2012-0234.  Research leading
to these results has received funding from the European Research
Council under the European Union’s Seventh Framework Programme
(FP7/2007-2013) including ERC grant agreements 240672, 291329, and
306478.

We are grateful for the extraordinary contributions of our CTIO
colleagues and the DECam Construction, Commissioning and Science
Verification teams in achieving the excellent instrument and telescope
conditions that have made this work possible.  The success of this
project also relies critically on the expertise and dedication of the
DES Data Management group.

The VISTA Hemisphere Survey (VHS) is based on observations obtained as part of  ESO Progamme 179.A-2010 (PI: McMahon).

This paper has gone through internal review by the DES collaboration.

\bibliographystyle{apj}
\bibliography{bib}{}

\appendix
\section{Affiliations}
\label{sec:affiliations}
$[1]$ Cerro Tololo Inter-American Observatory, National Optical Astronomy Observatory, Casilla 603, La Serena, Chile\\
$[2]$ Department of Physics \& Astronomy, University College London, Gower Street, London, WC1E 6BT, UK\\
$[3]$ Fermi National Accelerator Laboratory, P. O. Box 500, Batavia, IL 60510, USA\\
$[4]$ Department of Physics, ETH Zurich, Wolfgang-Pauli-Strasse 16, CH-8093 Zurich, Switzerland\\
$[5]$ Department of Astrophysical Sciences, Princeton University, Peyton Hall, Princeton, NJ 08544, USA\\
$[6]$ Institute of Cosmology \& Gravitation, University of Portsmouth, Portsmouth, PO1 3FX, UK\\
$[7]$ Kavli Institute for Cosmology, University of Cambridge, Madingley Road, Cambridge CB3 0HA, UK\\
$[8]$ Institute of Astronomy, University of Cambridge, Madingley Road, Cambridge CB3 0HA, UK\\
$[9]$ Institut de Ci\`encies de l'Espai, IEEC-CSIC, Campus UAB, Carrer de Can Magrans, s/n,  08193 Bellaterra, Barcelona, Spain\\
$[10]$ Department of Physics and Astronomy, University of Pennsylvania, Philadelphia, PA 19104, USA\\
$[11]$ Kavli Institute for Particle Astrophysics \& Cosmology, P. O. Box 2450, Stanford University, Stanford, CA 94305, USA\\
$[12]$ Sorbonne Universit\'es, UPMC Univ Paris 06, UMR 7095, Institut d'Astrophysique de Paris, F-75014, Paris, France\\
$[13]$ CNRS, UMR 7095, Institut d'Astrophysique de Paris, F-75014, Paris, France\\
$[14]$ Center for Cosmology and Astro-Particle Physics, The Ohio State University, Columbus, OH 43210, USA\\
$[15]$ Institut de F\'{\i}sica d'Altes Energies, Universitat Aut\`onoma de Barcelona, E-08193 Bellaterra, Barcelona, Spain\\
$[16]$ Jodrell Bank Centre for Astrophysics, School of Physics and Astronomy, University of Manchester, Oxford Road, Manchester, M13 9PL, UK\\
$[17]$ SLAC National Accelerator Laboratory, Menlo Park, CA 94025, USA\\
$[18]$ Observat\'orio Nacional, Rua Gal. Jos\'e Cristino 77, Rio de Janeiro, RJ - 20921-400, Brazil\\
$[19]$ Laborat\'orio Interinstitucional de e-Astronomia - LIneA, Rua Gal. Jos\'e Cristino 77, Rio de Janeiro, RJ - 20921-400, Brazil\\
$[20]$ National Center for Supercomputing Applications, 1205 West Clark St., Figure, IL 61801, USA\\
$[21]$ Department of Astronomy, University of Illinois, 1002 W. Green Street, Urbana, IL 61801, USA\\
$[22]$ Department of Physics, University of Michigan, Ann Arbor, MI 48109, USA\\
$[23]$ George P. and Cynthia Woods Mitchell Institute for Fundamental Physics and Astronomy, and Department of Physics and Astronomy, Texas A\&M University, College Station, TX 77843,  USA\\
$[24]$ Excellence Cluster Universe, Boltzmannstr.\ 2, 85748 Garching, Germany\\
$[25]$ Faculty of Physics, Ludwig-Maximilians University, Scheinerstr. 1, 81679 Munich, Germany\\
$[26]$ Universit\"ats-Sternwarte, Fakult\"at f\"ur Physik, Ludwig-Maximilians Universit\"at M\"unchen, Scheinerstr. 1, 81679 M\"unchen, Germany\\
$[27]$ Kavli Institute for Cosmological Physics, University of Chicago, Chicago, IL 60637, USA\\
$[28]$ Jet Propulsion Laboratory, California Institute of Technology, 4800 Oak Grove Dr., Pasadena, CA 91109, USA\\
$[29]$ Department of Astronomy, University of Michigan, Ann Arbor, MI 48109, USA\\
$[30]$ Max Planck Institute for Extraterrestrial Physics, Giessenbachstrasse, 85748 Garching, Germany\\
$[31]$ Centro de Investigaciones Energ\'eticas, Medioambientales y Tecnol\'ogicas (CIEMAT), Madrid, Spain\\
$[32]$ Department of Physics, The Ohio State University, Columbus, OH 43210, USA\\
$[33]$ Australian Astronomical Observatory, North Ryde, NSW 2113, Australia\\
$[34]$ Argonne National Laboratory, 9700 South Cass Avenue, Lemont, IL 60439, USA\\
$[35]$ Departamento de F\'{\i}sica Matem\'atica,  Instituto de F\'{\i}sica, Universidade de S\~ao Paulo,  CP 66318, CEP 05314-970 S\~ao Paulo, Brazil\\
$[36]$ Department of Astronomy, The Ohio State University, Columbus, OH 43210, USA\\
$[37]$ Instituci\'o Catalana de Recerca i Estudis Avan\c{c}ats, E-08010 Barcelona, Spain\\
$[38]$ Lawrence Berkeley National Laboratory, 1 Cyclotron Road, Berkeley, CA 94720, USA\\
$[39]$ Department of Physics and Astronomy, Pevensey Building, University of Sussex, Brighton, BN1 9QH, UK\\
$[40]$ Brookhaven National Laboratory, Bldg 510, Upton, NY 11973, USA\\
$[41]$ Department of Physics, University of Illinois, 1110 W. Green St., Urbana, IL 61801, USA\\
$[42]$ South East Physics Network, (www.sepnet.ac.uk), UK\\
$[43]$ Instituto de F\'{i}sica Te\'{o}rica IFT-UAM/CSIC, Universidad Aut\'{o}noma de Madrid, 28049 Madrid, Spain\\
$[44]$ Department of Astronomy, University of California, Berkeley,  501 Campbell Hall, Berkeley, CA 94720, USA\\
$[45]$ Instituto de F\'\i sica, UFRGS, Caixa Postal 15051, Porto Alegre, RS - 91501-970, Brazil\\
$[46]$ Department of Physics, University of Surrey, Guildford GU2 7XH, UK\\
$[47]$ Institut de F\'{\i}sica d'Altes Energies (IFAE), The Barcelona Institute of Science and Technology, Campus UAB, 08193 Bellaterra (Barcelona) Spain\\
$[48]$ School of Sciences, European University Cyprus, 6 Diogenis Str., Engomi, 1516 Nicosia, Cyprus\\
$[49]$ School of Mathematics and Physics, University of Queensland, QLD 4072, Australia\\
$[50]$ School of Physics and Astronomy, University of Southampton, Southampton, SO17 1BJ, UK\\
$[51]$ ICRA, Centro Brasileiro de Pesquisas F\'isicas, Rua Dr. Xavier Sigaud 150, CEP 22290-180, Rio de Janeiro, RJ, Brazil\\
$[52]$ Dipartimento di Fisica e Scienze della Terra, Universit\`a degli Studi di Ferrara, Via Saragat 1, I-44122 Ferrara, Italy\\
$[53]$ Department of Physics and Santa Cruz Institute for Particle Physics University of California, Santa Cruz, CA 95064, USA\\
$[54]$ Institute of Astronomy, ETH Zurich, 8093 Zurich, Switzerland\\
$[55]$ University of Nottingham, School of Physics \& Astronomy, Nottingham NG7 2RD\\
$[56]$ Wisconsin IceCube Particle Astrophysics Center (WIPAC), Madison, WI 53703, USA\\
$[57]$ Department of Physics, University of Wisconsin–Madison, Madison, WI 53706, USA\\
$[58]$ Department of Astronomy and Astrophysics, University of Chicago, 5640 South Ellis Avenue, Chicago, IL 60637\\
$[59]$ Department of Physics, University of Arizona, 1118 E 4th St, Tucson, AZ 85721\\
$[60]$ National Center for Supercomputing Applications, 1205 West Clark St, Urbana, IL 61801, USA

\end{document}